# Spatiotemporal patterns of Io's bright transient eruptions, 1978-2022

Christian D. Tate[1], Julie A. Rathbun[1,2], Alexander G. Hayes[1], Rosaly M. C. Lopes[3], Madeline Pettine[1]

1. Department of Astronomy, Cornell University, Ithaca, NY, USA
2. Planetary Science Institute, Tucson, AZ, USA
3. Jet Propulsion Laboratory, California Institute of Technology, Pasadena, CA, USA

## Abstract

This study analyzes Io's thermally detected volcanic outbursts and mini-outbursts, generally called bright transient eruptions. We examine their evolving characteristics over the history of outburst observations between the Voyager flybys in 1978 and 2022. We catalog, compare, and interpret the data of these bright transient eruptions from several spacecraft flybys and numerous ground-based observation campaigns. To test the spatiotemporal behavior of these events, we compare them to a population of randomly spaced, stochastic events with an equal likelihood of occurrence anywhere on Io's surface. We find that the aggregate of all outbursts is consistent with a random distribution across Io, whereas mini-outbursts strongly prefer the trailing hemisphere (180 to 360 W). On shorter timescales, however, outbursts show a significant change in spatiotemporal behavior before and after the year 2012. Outbursts from 1995 to 2007 favor the northern leading hemisphere, while outbursts from 2013 to 2021 favor the southern trailing hemisphere. These temporally separated clusters of outbursts are remarkably similar to Io's two primary mountainous regions, indicating that outbursts may be related to mountain-forming activity. These trends show how bright transient eruptions are distinct from Io's other forms of volcanism. These could be essential constraints to assess models of Io's interior heat transport between tidal generation and volcanic distribution.

## 1 Introduction

Jupiter's innermost Galilean moon, Io, is the most volcanically active object in the solar system. With far more energy to release than Earth and other bodies and less surface area to dissipate it, Io is in a class of its own. Its most intense volcanic eruptions, called outbursts, occur about monthly (Spencer & Schneider, 1996), whereas Earth's subaerial "large eruptions" (that produce ≥0.1 km$^3$ tephra) occur once or twice yearly (Siebert et al., 2015). Because Io's geological clock runs several orders of magnitude faster than Earth's, Io's liveliness lets us observe volcanism at both larger spatial scales and on faster timescales.

Although Io's current data are relatively coarse in spatiotemporal resolutions, the persistent inquiry of scientists and advances in observational techniques over the last fifty years have produced a compelling, complex, and exciting portrait of Io. Nevertheless, a theoretical understanding of Io's volcanic mechanisms remains elusive. While other studies have summarized the significant achievements and limitations of the current knowledge about Io (de Kleer & Rathbun, 2023; McEwen et al., 2023), it appears that much more data and analysis are necessary to understand what makes Io tick. It is difficult to characterize Io's steady state behavior and to constrain its stochastic and periodic departures from that hypothetical steady state. We explore these gaps in knowledge by studying the historical record of Io's bright transient eruptions and how their behavior has changed over the last fifty years.

Previous studies have analyzed portions of Io’s outburst dataset. Veeder et al. (2012) examined 14 of the first outbursts observed by the Voyager and Galileo spacecraft and ground-based infrared photometry starting in the late 1970s. This study paid close attention to the Galileo dataset, which remains the source of Io's best spatial and spectral resolution even today. While Veeder et al. (2012) examined outbursts up to the year 2001 from the perspective of Io's total heat budget, Cantrall et al. (2018) investigated the geological context of 5 outbursts and 17 mini-outbursts that occurred from 2001 to 2015. More recently, de Kleer and Rathbun (2023) compiled 33 outburst and “sub-outburst” events from 1978 to 2018, and a recent IRTF observational campaign found seven new outbursts between 2017 and 2021 (Tate et al., 2023). Our review identifies 66 likely outburst and mini-outburst observed between 1978 and 2022. This list includes every event with some evidence of it being in the class of a bright transient eruption. We carefully filter the less confident events from this inclusive list and apply our statistical analyses to a more robust subset of events.

# 2 Data

Campaigns using NASA's Infrared Telescope Facility (IRTF) started in the 1990s and remain active even after next-generation adaptive optics (AO) telescope systems came online in the late 1990s. These IRTF observation campaigns of Io's volcanism supply valuable long-term data with updated instruments over the course of three decades (Rathbun et al. 2010; Tate et al., 2023). In contrast to earlier observations, many of the outburst detections made after the year 2000 were primarily made with adaptive optics system (AO) telescopes including Keck, Gemini, and the European Southern Observatory (Marchis et al., 2002; Marchis et al., 2004; de Pater et al., 2016a; de Pater et al., 2016b; de Kleer et al. 2019b). It is difficult to overestimate the importance of ground-based techniques in the years following the Galileo mission. Nearly two decades of consistent, high-resolution surveys of Io have created an amazing dataset with which to study the character of Io's volcanism over decadal timescales.

We gather all published data about large-scale infrared observation campaigns capable of detecting outbursts on Io. The number of events used in our various analyses are listed in Table 1, and Table 2 is the inclusive list of Io's bright transient eruptions. The major campaigns and spacecraft missions are summarized in Table 3. Of the 66 events, 30 are outbursts, and 36 are mini-outbursts. The rationale for classifying these events is explained below. Although the instruments and techniques of these campaigns vary greatly, we compile plausible outbursts and mini-outbursts discovered in the major infrared detection campaigns. We generally accept the designation of an outburst or mini-outburst if the primary reference and subsequent publications give compelling evidence. For consistency across the whole dataset, however, we also devise and filter by a more uniform classification metric.

## 2.1 Criteria used for including outbursts and mini-outbursts in this analysis

This study aims to characterize the spatial, temporal, and thermal properties of bright transient events (e.g., outbursts and mini-outbursts). We use the definitions of outburst and mini-outbursts described by Tate et al. (2023), which require evidence that the eruption is sufficiently bright and confined to a small region on Io. Although not considered here, outbursts can also be inferable from enhancements in Io's plasma torus (Brown & Bouchez, 1997; Morgenthaler et al., 2019 and 2022). Even for direct detections, however, the use of the term "volcanic outburst" is not entirely consistent in the Io literature (see Tate et al., 2023 for a discussion of this point). To clarify and standardize the criteria necessary for our analyses of outbursts or mini-outbursts, we identify three primary criteria:

1. **A bright transient eruption event has a large infrared output.** We use the threshold in the intensity in the 3.8 μm Lp-band, for which an outburst emits $I_{3.8\mu m}$ > 150 GW/sr/μm and a mini-outburst emits $I_{3.8\mu m}$ > 30 GW/sr/μm (de Kleer & de Pater, 2016a; Tate et al., 2023). Note that this intensity should be corrected for reflected sunlight, emission angle, and other photometric effects.
2. **An event is spatially constrained** if it is confined to a small region on Io. We adopt a maximum localization uncertainty of 15 degrees or ~500 km in north-south and west-east directions. This localization threshold is near the best achievable with NASA's IRTF.
3. **An event is thermally constrained** if several 1-10 μm intensity measurements can constrain its effective temperature, area, and total power. Four or more bands are preferred if the event is thoroughly non-uniform and is not well characterized with a 1-temperature fit.

For our study, an outburst or mini-outburst is confirmed and spatially characterized if its data satisfies the first and second criteria. An event is thermally characterized if it satisfies the first and third criteria (assuming that care is taken to mitigate spectral contamination from other hot spots and reflected sunlight). Finally, an event is fully characterized if it satisfies all three criteria. Further constraints are possible, such as time-resolved changes in intensity and modeling the lava flow and cooling rates (Davies et al., 2000, 2005, 2006, 2010, 2014).

When the 3.8 μm intensity is not directly measured, criterion (1) can be satisfied by interpolating or extrapolating several wavelength measurements to 3.8 μm or by evaluating the 3.8 μm of a black-body fit. This is common for ground-based and New Horizons instruments that take 1.0-2.2 μm spectra (Marchis et al., 2002; Tsang 2014; Spencer et al., 2007). Intensities at wavelengths less than 1 micron, such as the <1-micron wavelength filters of the Hubble Space Telescope, have not by themselves been successful at constraining Io's eruption temperatures (Marchis et al., 2002; Milazzo et al., 2005). In the case of noisy intensity measurements with abnormally large uncertainties, such as full-disk sunlit imaging with the IRTF (Tate et al., 2023), the lower limit of the peak intensity should exceed the outburst or mini-outburst threshold.

Not all confirmed outbursts need to satisfy criterion (2) if they have a sufficiently high 3.8 μm intensities, often more than twice the intensity threshold $I_{3.8\mu m}$ > 300 GW/sr/μm. This exception is essential for early disk-integrated techniques and the sunlit IRTF campaign (Tate et al., 2023). Several early detections fall into this category, such as the colossal 1978 and 1986 outbursts (Witteborn et al., 1979; Veeder et al., 1994; Blaney et al., 1995), neither of which are named or localized. Some confirmed outbursts are thermally constrained if they have high-quality multi-spectral intensity measurements that control for spectral contamination from reflected sunlight and other hot spots. For instance, we consider 202010A a thermally (but not spatially) constrained outburst because of its high-quality multi-spectral intensity measurements were taken in Jovian eclipse and occultation, which removes contamination from reflected sunlight and isolates the hot spot along one direction (Tate et al., 2023). For these reasons, the collections of spatially and thermally characterized events given in Tables 1 and 2 are overlapping but not identical.

| | Outbursts | Mini-outbursts | Both |
|---|---|---|---|
| Total number of events | 30 | 36 | 66 |
| Localized events | 17 | 27 | 44 |
| Loc. 1995-2011 | 10 | 14 | 24 |
| Loc. 2013-2022 | 7 | 13 | 20 |
| Unique locations | 13 | 16 | 29 |
| Events at Tvashtar Paetra | 6 | 0 | 6 |
| Events at Pillan Patera | 2 | 4 | 6 |
| Events at Loki Patera | 0 | >10 | >10 |
| Thermally constrained events | 18 | 16 | 34 |
| Therm. and loc. constrained | 13 | 15 | 28 |
| Median effective temperature | 1240 K | 905 K | 1150 K |
| Median effective total power | 7.8 TW | 1.2 TW | 3.7 TW |

**Table 1.** The total number of large transient volcanic events used in this study.

| Date | Location Name *a* | Type | Latitude North | Longitude West | Spatially const. *b* | Thermally const. *b* | Wavelengths [μm] | Intensity *c* | Effective Temperature [K] | Effective Power [TW] | Notes |
|---|---|---|---|---|---|---|---|---|---|---|---|
| 1978/01/26 | – | outburst | – | 23 ± 90 | no | no | 2–4 | – | – | – | *d* |
| 1978/02/20 | – | outburst | – | 60 ± 78 | no | no | 1.2–5.4 | – | 600 ± | *17.0* | *e* |
| 1979/06/11 | Surt | outburst | 28 ± 20 | 39 ± 86 | no | no | 5 | – | 600 ± | *28.0* | *f* |
| 1986/08/07 | – | outburst | – | 69 ± 71 | no | yes | 4.8 – 8.7 | – | 1550 ± | 58.0 | *g* |
| 1990/01/09 | Loki Patera? | outburst | – | 310 ± 51 | no | yes | 4.8 – 8.7 | – | 1225 – 1600 | 11.0 | *g* |
| 1995/03/02 | 9503A, Arusha? | outburst | -45 ± 15 | 95 ± 15 | yes | no | 3.5 – 4.8 | 650 ± 150 | 600 ± | 3.6 | *h* |
| 1995/09/27 | 9509A | outburst | -45 ± 45 | 77 ± 45 | no | no | 3.8 | 1400 ± 200 | – | – | *h* |
| 1996/10/06 | 9610A | outburst | 75 ± 15 | 35 ± 15 | yes | yes | 1.7 – 4.8 | 360 ± 30 | >1400 | 3.9 | *h* |
| 1996/12/19 | Marduk Fluctus | mini-outb. | -28.4 ± 2 | 210 ± 2 | yes | yes | 1.8 - 5.2 | 80 ± 0 | 1144 ± | 1.1 | *i* |
| 1997/06/28 | Pillan Patera | outburst | -12 ± 2 | 244 ± 2 | yes | yes | 0.4 - 5.2 | – | 1600 ± | 3.4 | *j* |
| 1999/06/22 | 9906A | outburst | -10 ± 20 | 14 ± 20 | no | no | 3.5 | 180 ± 0 | – | – | *k* |
| 1999/08/02 | 9908A, Gish Bar? | outburst | 14 ± 12 | 74 ± 12 | yes | yes | 2.3 - 4.7 | 1805 ± 0 | 1250 ± | 27.0 | *k* |
| 1999/11/25 | Loki Patera | mini-outb. | 13 ± 5 | 309 ± 5 | yes | no | 3.5 - 4.8 | 127 ± 8 | – | – | *k* |
| 1999/11/25 | Pele Patera | mini-outb. | -19 ± 5 | 255 ± 5 | yes | no | 3.5 - 4.8 | 49 ± 2 | | | *k* |
| 1999/11/26 | Tvashtar Catena | outburst | 62 ± 2 | 120 ± 2 | yes | yes | 3.5 - 4.8 | 885 ± 0 | 1300 ± | 24.0 | *k* |
| 2000/02/20 | 0002A | outburst | 64 ± 20 | 80 ± 4 | no | no | 3.5 | 200 ± 25 | – | – | *k* |
| 2000/12/16 | Tvashtar Catena | outburst | 62 ± 2 | 120 ± 2 | no | no | visible - 0.98 | – | – | – | *l* |
| 2001/02/19 | Tvashtar Catena | outburst | 59 ± 4 | 124 ± 3 | yes | yes | 1.6 – 2.2 | – | 810 ± | 11.0 | *m* |
| 2001/02/19 | North Amirani | outburst | 27 ± 3 | 118 ± 3 | yes | yes | 1.6 – 2.2 | – | 990 ± | 13.8 | *m* |
| 2001/02/22 | Surt | outburst | 41 ± 4 | 338 ± 5 | yes | yes | 1.2 – 2.2 | – | 1240 ± | 78.0 | *m* |
| 2001/08/05 | Dazhbog Patera | mini-outb. | 54 ± 4 | 302 ± 5 | no | no | 1 – 100 | – | – | – | *n* |
| 2001/08/06 | Thor | mini-outb. | 26 ± 1 | 147 ± 1 | no | no | 0.7 – 5.2 | 16 ± 0 | – | – | *o* |
| 2002/02/21 | Loki Patera | mini-outb. | 13 ± 2 | 309 ± 4 | yes | no | 3.4 - 4.8 | 55 ± 9 | 426 ± – | 4.7 | *p* |
| 2004/05/30 | Sui Jen Patera | mini-outb. | -19.2 ± 1 | 4 ± 1 | yes | yes | Jcont – 4.7 | 59 ± 9 | 884 ± | 1.2 | *q* |
| 2004/05/30 | Tung Yo Patera | mini-outb. | -18.3 ± 1 | 0.9 ± 2 | yes | yes | 3.8 – 4.7 | 72 ± 12 | 550 ± | 2.5 | *q* |
| 2005/05/31 | South of Babbar | mini-outb. | -50 ± 2 | 278 ± 4 | yes | yes | 2.2 – 4.7 | 57 ± 14 | 925 ± | 1.2 | *q* |
| 2006/04/17 | Tvashtar Catena | outburst | 59 ± 1 | 121 ± 1 | no | no | 1.0 – 2.4 | – | – | – | *r* |
| 2006/06/02 | Tvashtar Catena | outburst | 59 ± 1 | 121 ± 1 | yes | yes | 1.0 – 2.4 | – | 1240 ± 4 | 7.7 | *r* |
| 2007/03/01 | East Girru | mini-outb. | 21 ± | 236 ± | no | no | 0.4 – 2.1 | – | 1017 ± | | *s* |
| 2007/03/01 | Tvashtar Catena | outburst | 59 ± 1 | 121 ± 1 | yes | yes | 0.4 – 2.1 | – | 1239 ± | 4.9 | *s* |
| 2007/08/14 | Pillan Patera | mini-outb. | -8.5 ± 1 | 245 ± 1 | yes | yes | 2.2 – 4.8 | 30 ± 5 | 840 ± 40 | 0.5 | *t* |
| 2008/07/24 | Pillan Patera | mini-outb. | -7 ± 2 | 244 ± 1 | yes | no | 4.0 | 157 ± 24 | – | – | *t* |
| 2009/09/10 | Loki Patera | mini-outb. | 13 ± 2 | 309 ± 4 | yes | no | 2.2 - 4.8 | 75 ± 12 | 488 ± – | 3.6 | *p* |
| 2010/06/28 | Pillan Patera | mini-outb. | -8.5 ± 1 | 245 ± 1 | yes | yes | 3.5 - 4.8 | 30 ± 5 | 640 ± 50 | 0.5 | *t* |
| 2010/09/17 | Kanehekili Fluctus | mini-outb. | -18.5 ± 1.5 | 32 ± 2.2 | yes | no | 3.8 – 4.8 | 38 ± 6 | 335 ± – | 15.5 | *u* |
| 2010/09/20 | Loki Patera | mini-outb. | 13 ± 2 | 309 ± 4 | yes | no | 3.8 – 4.8 | 97 ± 15 | 486 ± – | 4.8 | *p* |
| 2010/11/23 | Kanehekili Fluctus | mini-outb. | -18.5 ± 1.5 | 32 ± 2.2 | yes | yes | 3.8 – 4.8 | 31 ± 5 | 525 ± | 1.2 | *u* |
| 2013/08/15 | Heno Patera | outburst | -56 ± 3 | 307 ± 3 | yes | yes | 2.3 – 4.1 | 270 ± 70 | 719 ± 3 | 5.0 | *v* |
| 2013/08/20 | Loki Patera | mini-outb. | 13 ± 2 | 309 ± 4 | yes | no | 2.2 - 4.8 | 325 ± 80 | 514 ± – | 5.8 | *p* |
| 2013/08/20 | Rarog Patera | outburst | -39 ± 2 | 306 ± 3 | yes | yes | 2.3 – 3.8 | 136 ± 20 | 1041 ± 2 | 7.8 | *v* |
| 2013/08/29 | 201308C | outburst | 29 ± 2 | 227 ± 6 | yes | yes | 3.8 | > 500 | 1,900 | 24.5 | *w* |
| 2014/10/22 | Chors Patera | mini-outb. | 65 ± 2 | 247 ± 4 | yes | no | 3.8 | 57 ± 19 | 398 ± 4 | 3.1 | *x* |
| 2014/10/31 | Loki Patera | mini-outb. | 13 ± 2 | 307 ± 4 | no | no | 3.8, 4.7 | 78 ± 20 | 416 ± 27 | 7.9 | *w* |
| 2015/01/10 | Mithra Patera | mini-outb. | -59 ± 2 | 266 ± 4 | yes | no | 3.8 | 55 ± 12 | – ± – | – | *x* |
| 2015/01/26 | Kurdalagon Patera | mini-outb. | -49 ± 1 | 219 ± 2 | yes | yes | 2.2, 3.8 | 56 ± 9 | 1184 ± | 1.0 | *x* |
| 2015/02/18 | Pillan Patera | mini-outb. | -11.5 ± 1 | 243 ± 1 | no | no | 3.8 | 80 ± 16 | ± | | *t* |
| 2015/04/01 | Sethlaus, Gabija | mini-outb. | -50 ± 1 | 198 ± 2 | yes | yes | 3.0 – 4.7 | 33 ± 5 | 616 ± 42 | 0.7 | *x* |
| 2015/04/05 | Kurdalagon Patera | mini-outb. | -49 ± 2 | 224 ± 2 | yes | yes | 2.2, 3.8 | 68 ± 11 | 1290 ± | 1.3 | *x* |
| 2015/12/25 | Amaterasu Patera | mini-outb. | 39 ± 1 | 304 ± 1 | yes | yes | 3.0 – 4.7 | 43 ± 6 | 412 ± | 5.3 | *x* |
| 2016/01/22 | Loki Patera | mini-outb. | 13 ± 2 | 309 ± 4 | yes | no | 2.2 - 4.8 | 147 ± 22 | 505 ± – | 6.5 | *p* |
| 2016/06/20 | Shamash Pater | mini-outb. | -33 ± 2 | 151 ± 2 | yes | yes | 2.2, 3.8 | 53 ± 9 | 1000 ± | *0.9* | *x* |
| 2016/06/27 | Illyrikon Regio | mini-outb. | -71 ± 2 | 180 ± 2 | yes | yes | 2.2, 3.8 | 125 ± 69 | 1210 ± | *2.2* | *x* |
| 2017/03/17 | P95 | mini-outb. | -10 ± 2 | 128 ± 2 | yes | yes | 2.2, 3.8 | 58 ± 13 | 1020 ± | *1.0* | *x* |
| 2017/06/25 | Loki Patera | mini-outb. | 13 ± 2 | 308 ± 4 | no | no | 2.2 - 4.8 | 140 ± 20 | – | – | *y* |
| 2018/01/18 | 201801A | outburst | 10 ± 14 | 220 ± 15 | no | no | 3.8 | 193 ± 48 | – | – | *z* |
| 2018/05/10 | UP 254W | outburst | -37 ± 2 | 252 ± 1 | yes | yes | 2.2, 3.8 | 325 ± 117 | 960 ± | 2.1 | *aa* |
| 2018/05/27 | Isum Patera | mini-outb. | 31 ± 2 | 205 ± 2 | yes | yes | 3.1 – 4.7 | 64 ± 16 | 1200 ± | *1.1* | *x* |
| 2019/05/08 | 201905A, Acala? | outburst | 8 ± 9 | 332 ± 11 | no | yes | 2.3 – 4.8 | 298 ± 24 | 1160 ± 52 | 5.1 | *bb* |
| 2019/06/25 | 201906A, Acala | outburst | 8 ± 9 | 332 ± 11 | yes | yes | 2.3 – 4.8 | 241 ± 15 | 1214 ± 40 | 5.4 | *bb* |
| 2020/02/15 | Loki Patera | mini-outb. | 13 ± 2 | 308 ± 4 | no | no | 2.2 - 4.8 | 120 ± 20 | – | – | *y* |
| 2020/02/17 | Laki-oi | mini-outb. | -42 ± 5 | 55 ± 5 | no | no | 4.8 | 61 ± 0 | – | – | *cc* |
| 2020/11/15 | 202010A | outburst | 5 ± 60 | 20 ± 16 | no | yes | 1.6 – 4.8 | 309 ± 33 | 1279 ± 35 | 5.8 | *z* |
| 2021/06/23 | Loki Patera | mini-outb. | 13 ± 2 | 308 ± 4 | no | no | 2.2 - 4.8 | 90 ± 20 | – | – | *y* |
| 2021/08/13 | Pillan Patera | outburst | -9 ± 9 | 242 ± 10 | yes | no | 3.8, 4.8 | 169 ± 36 | 500 ± – | *6.4* | *z* |
| 2021/08/27 | 202108D | outburst | -10 ± 10 | 183 ± 13 | yes | no | 3.8, 4.8 | 395 ± 51 | 500 ± – | *14.0* | *z* |
| 2022/11/15 | Kanehekili Fluctus | mini-outb. | -18.5 ± 1.5 | 32 ± 2.2 | yes | yes | 1.7-5.3 | 35 ± 1 | 525 ± | 1.2 | *dd* |

**Table 2.** List the known, thermally-detected outbursts and mini-outbursts from 1978 to 2022.
Table footnotes:

a. The location name is the hot spot name or an alphanumeric designation assigned in its first publication. A hot spot name follows some alphanumeric designations for several less certain events.
b. The criteria for spatial and thermal constraints are given in section 2.1
c. The intensity is in units of GW/sr/micron and nominally at the 3.8 micron or the Lp-band. When 3.8-micron intensity was not measured, the event's footnote gives the used wavelength.
d. Spencer and Schneider (1996); and unpublished data from Fink, Lebofsky, Larson.
e. Witteborn et al. (1979); Spencer and Schneider (1996).
f. Sinton et al., (1980), Spencer and Schneider (1996).
g. Veeder et al. (1994), Blaney et al. (1995), Spencer and Schneider (1996).
h. Spencer and Schneider (1996), Spencer et al. (1997), Veeder et al. (2012). Intensities are at 3.5 μm. The 9509A event was not included in the Veeder et al. (2012) list of outbursts.
i. Davies et al. (2018).
j. Davies et al. (2001), Veeder et al. (2012).
k. Howell et al. (2001). For the 1999 Tvashtar event, also see Marchis et al. 2002; Milazzo et al., (2005); Davies et al. (2010). The 9906A and 0002A events were not included in the Veeder et al. (2012) list of outbursts. Veeder et al. (2012) located 9908A at 85W.
l. Milazzo et al., (2005), Davies et al. (2010). This event is not used in the location statistics because it was likely part of 2001/02/19 eruption.
m. Marchis et al. (2002). Tvashtar was also seen two months earlier on 2000/12/15. Io was highly active in 2001. Surt erupted in the most powerful outburst on record.
n. Rathbun et al. (2004), Marchis et al. (2005).
o. Lopes et al. (2004).
p. de Pater et al. (2017) and de Kleer and de Pater (2016a) for events after 2013.
q. de Pater et al. (2016b), Marchis et al. (2004). Some values differ from those reported in Cantrall et al. (2018).
r. Laver et al. (2007), Cantrall et al. (2018). The 2006/04/17 measurement was saturated but comparable to the 2006/06/02 outburst.
s. Tsang et al. (2014), Spencer et al., (2007). On 2007 January 18 IRTF observed a bright feature near Tvashtar (Rathbun & Spencer, 2009), and a short time later on 2007 March 1 the New Horizons blyby observed an outburst and a Pele-type plume at Tvashtar.
t. de Pater et al. (2016a) reported mini-outbursts at Pillan in 2007, 2010, and 2015. Lellouch et al. (2015) detected another mini-outburst in 2008 at 4 micron intensity. Pillan was also active on 2015/03/31. The 2010 Pillan event is missing from Cantrall et al. (2018).
u. de Pater et al. (2016a) detected two mini-outbursts at Kanehekili Fluctus in 2010. A similar detection on 2010/08/21 comes close to a mini-outburst. Galileo saw two Prometheus-style plumes in 1997/5/6 and 1997/11/8.
v. de Pater et al. (2014a), de Kleer and de Pater (2016a), de Kleer et al (2019b). Heno qualified as a mini-outburst on 2013/08/20, 22, and 29. Rarog was also outbursting on 2013/08/22 and mini-outbursting 2013/08/15 to 2013/09/13.
w. de Kleer et al. (2014), de Kleer and de Pater (2016a) de Kleer et al (2019b). Juno did not detect a hot spot at the 201308C (JR159: 28.9 N, 228.8 W) location on orbit 10 (2017-12-16) but did on orbits 25 (2020-02-17) and (2021-02-21).

x. de Kleer and de Pater (2016a), Cantrall et al. (2018), de Kleer et al (2019b). Chors Patera was mini-outburst two days after its maximum intensity on 2014/10/22. Kurdalagon also approached mini-outburst levels on 2015/04/17.

y. de Kleer and Rathbun (2023), Tate et al. (2023).

z. Tate et al. (2023). 201801A (JR143: 20.3, 217 W) was detected on Juno's orbits 16 (2018-10-29), 24 (2019-12-26), 25 (2020-02-17), 32 (2021-02-21), and 33 (2021-04-15). It was not detected on orbits 10 (2017-12-16), 20 (2019-05-29), or 27 (2020-06-02). Pillan Patera (JR079) was detected on Juno's orbits 16 (2018-10-29), 17 (2018-12-21), and (2019-02-12). It was not detected on orbits 27 (2020-06-02) or 37 (2021-10-16). 202108D (JR102: 6.1 S, 186.4 W) was detected on Juno's orbits 25 (2020-02-17) and 32 (2021-02-21). It was not detected on orbits 33 (2021-04-15) or 37 (2021-10-16).

aa. UP 254W was reclassified as an outburst. The max 3.8-micron intensity comes from Tate et al. (2023), and the effective temperature and power come from de Kleer et al. (2019).

bb. Tate et al. (2023). IRTF localized 201906A within the Acala Fluces area. Acala (JR121: 8.3, 334.5 W) was detected on Juno's orbits 18 (2019-02-12), 20 (2019-05-29), 26 (2020-04-10) and 32 (2021-02-21). It was not detected on orbits 24 (2019-12-26) or 27 (2020-06-02).

cc. Zambon et al. (2023), Patine et al., (2023). Intensity is at 4.7 microns. Laki-oi (JR205) was detected on Juno's orbits 10 (2017-12-16), 11 (2018-02-07), 25 (2020-02-17). It was not detected on orbits 20 (2019-05-29) or 37 (2021-10-16).

dd. de Pater et al. (2023). Kanehekili (JR080) was detected on Juno's orbits 20 (2019-05-29), 24 (2019-12-26), and 25 (2020-02-17). It was not detected on orbit 37 (2021-10-16).

| Years | Observatory | Observations | Outbursts | Frequency | Mini-outbursts | Frequency | References |
|---|---|---|---|---|---|---|---|
| 1978 | LPL | 14 | 1 | 7.1% | – | – | Witteborn et al. (1979) |
| 1979 | Voyagers | 2 | 0 | 0.0% | – | – | McEwen and Soderblom, (1983) |
| 1979 – 1981 | IRTF, UKIRT, UH | 27 | 1 | 3.7% | – | – | Sinton et al. (1983) |
| 1982 – 1983 | UKIRT | 42 | 0 | 0.0% | – | – | Tittemore & Sinton (1989) |
| 1983 – 1993 | IRTF | 55 | 2 | 3.6% | – | – | Veeder et al. (1994), Blaney et al. (1995) |
| 1989 – 1992 | WIRO | 96 | 0 | 0.0% | – | – | Howell & Klassen (1995) |
| 1995 – 1997 | IRTF, Lowell | 56 | 3 | 5.4% | – | – | Spencer et al. (1997) |
| 1996 – 2001 | Galileo | 7 | 1 | 14% | 2 | 28.6% | Rathbun et al (2004), Lopes et al. (2004) |
| 1999 – 2000 | IRTF, WIRO, ESO | 28 | 4 | 14.3% | – | – | Howell et al. (2001) |
| 2001– 2007 | IRTF, WIRO | 33 | 0 | 0.0% | – | – | Rathbun & Spencer (2010) |
| 2001 | Keck, ESO | 13 | 3 | 23.1% | 0 | 0.0% | Marchis et al. (2002), (2005) |
| 2003 – 2007 | Keck | 12 | 2 | 16.7% | 6 | 50.0% | de Pater (2014a), (2014b), (2017) |
| 2007 | New Horizons | 2 | 1 | 50% | 1 | 50.0% | Laver et al. (2007), Tsang et al. (2014) |
| 2008 – 2009 | Keck | 11 | 0 | 0.0% | 2 | 18.2% | de Pater (2014a), (2014b), (2017) |
| 2010 – 2012 | Keck, Gemini | 18 | 0 | 0.0% | 4 | 22.2% | de Pater (2014b), (2017) |
| 2013 – 2018 | Keck, Gemini | 271 | 3 | 1.1% | 13 | 4.8% | de Kleer et al. (2019a) |
| 2016 – 2022 | IRTF (eclipse) | 77 | 3 | 3.9% | – | – | Tate et al. (2023) |
| 2016 – 2022 | IRTF (sunlit) | 99 | 4 | 4.0% | – | – | Tate et al. (2023) |
| 2017 – 2022 | Juno | 20 | 0 | 0.0% | 1 | 5.0% | Pettine et al. *(2023)* |
| 2022 | Keck, Gemini | 4 | 0 | 0.0% | 0 | 0.0% | de Pater et al., (2023) |
| 2022 | JWST | 1 | 0 | 0.0% | 1 | 100.0% | de Pater et al., (2023) |
| **All: 1978 – 2022** | **–** | **888** | **28** | **6.2 ± 1.2%** | **30** | **17 ± 3%** | |

**Table 3.** List of the major Io observation campaigns that are sensitive to the infrared signature of outbursts. Only the high sensitivity campaigns are used to measure the frequency of mini-outbursts.

| Years | Observations | Outburst events | Outburst frequency | Mini-outburst events | Mini-outburst frequency | Time-averaged total effective power *a* |
|---|---|---|---|---|---|---|
| **1978 – 1993** | **236** | **4** | **3.4 ± 1.7%** | **–** | **–** | **–** |
| **1995 – 2012** | **180** | **14** | **16 ± 4%** | **15** | **48 ± 12%** | **1.78 ± 0.47 TW** |
| **2013 – 2022** | **472** | **10** | **4.3 ± 1.4%** | **15** | **10 ± 3%** | **0.45 ± 0.14 TW** |
| **All, 1978 – 2022** | **888** | **28** | **6.2 ± 1.2%** | **30** | **17 ± 3%** | **0.69 ± 0.13 TW** |

**Table 4.** Summary of the combined Io observation campaigns sensitive to the infrared signature of outbursts and mini-outbursts. Note that the activity in 1995-2012 was about four times higher than before or after.

a. This estimate of time-average total effective power uses the median total power values in Table 1 and Figure 4. This value is the combined thermal output for both outbursts and mini-outbursts, of which the former emits 2-3 times more thermal power than the latter. This calculation assumes that events happen at the above-estimated frequencies and emit their median peak total effective power for 24 hours (given in Table 1 and Figure 4).

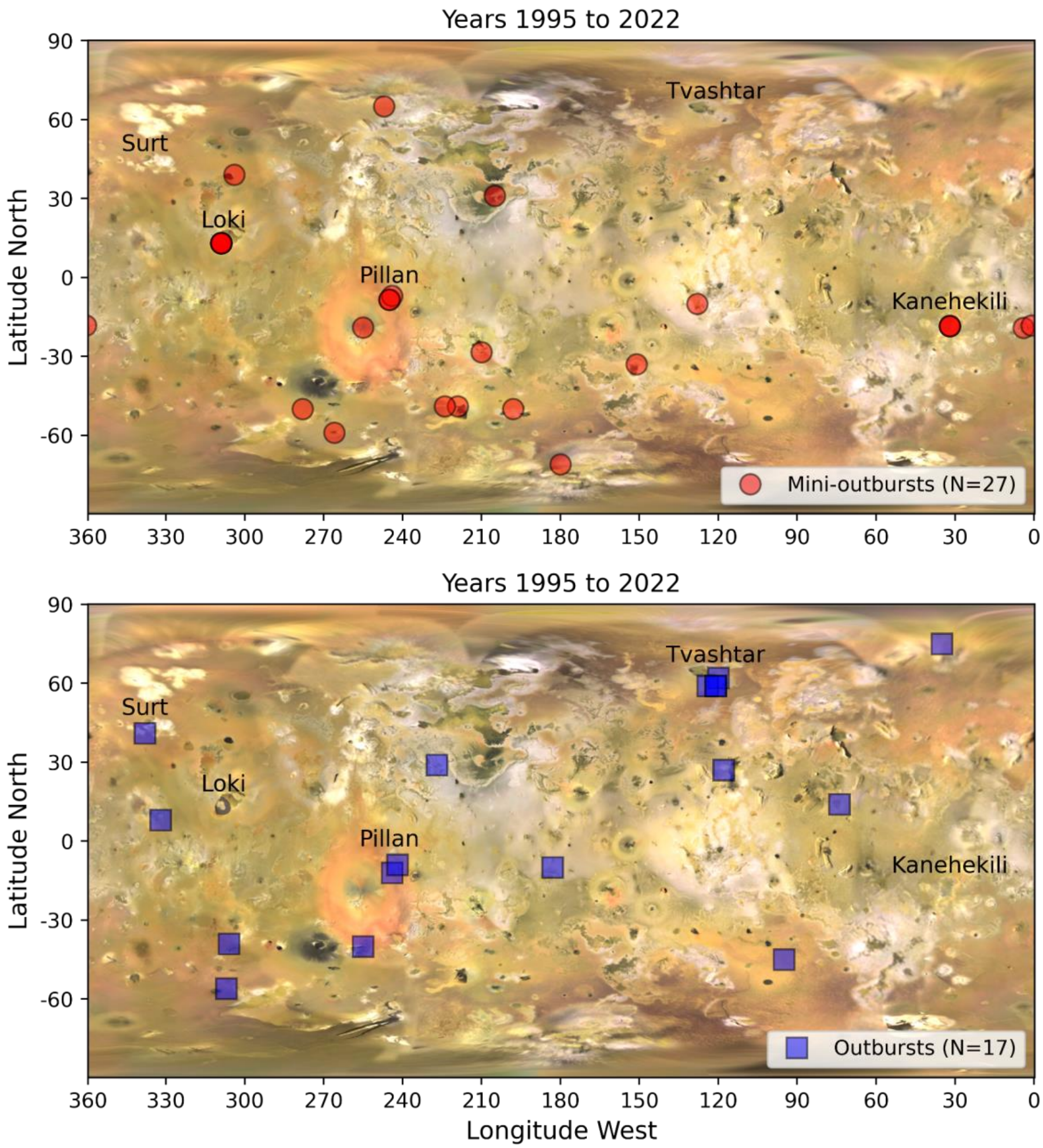

**Figure 1.** Mini-outbursts (top in red) and outbursts (bottom in blue) for the spatially constant events from 1995 to 2022. The global mosaic of Io is from Becker and Geissler (2005) and Williams et al. (2011) made from Galileo's high-resolution visible-NIR images. Note how mini-outbursts erupt primarily between longitudes 180W and 320W, outside of which they are confined to equatorial region. Outbursts by contrast appear more uniform.

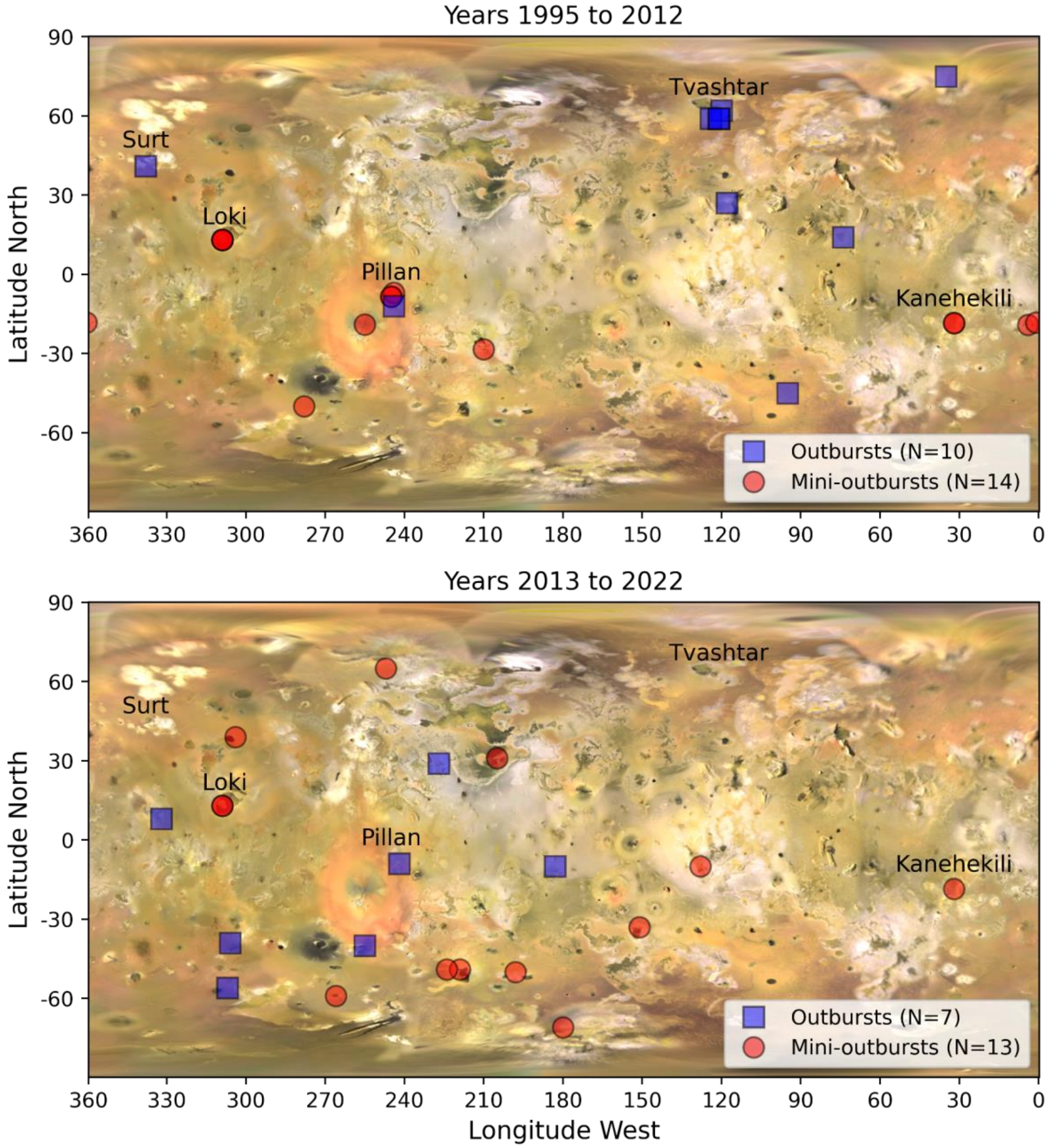

**Figure 2.** Same as Figure 1, except that outburst locations (blue) and mini-outburst locations (red) are in two time periods: 1995-2012 (top) and 2013-2022 (bottom). Note how outburst eruptions dramatically changed their global patterns after 2012.

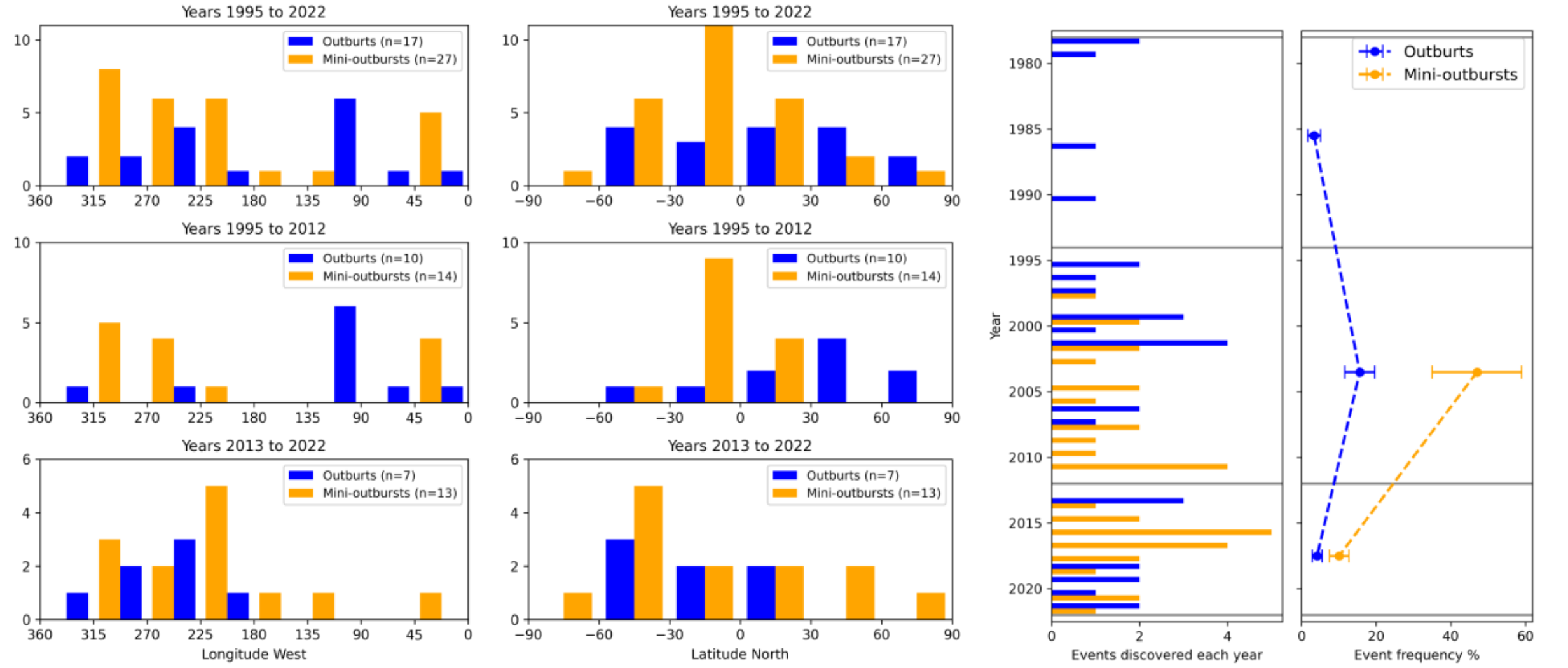

**Figure 3**: Histograms of the spatially constrained outbursts (blue) and mini-outbursts (orange) binned in longitude (left) and latitude (middle). The top row is for the years 1995-2022, which spans the dataset of localized events. The middle row is 1995-2012, and the bottom is 2013-2022. The right plots show the number of outbursts and mini-outbursts discovered each year from 1978 to 2022 (middle right) and the estimated frequency percentage of event occurrence for three periods (right). Note that the event detections include some that are not spatially constrained. See Table 5 for spatiotemporal statistics and Table 4 for frequency constraints for these periods.

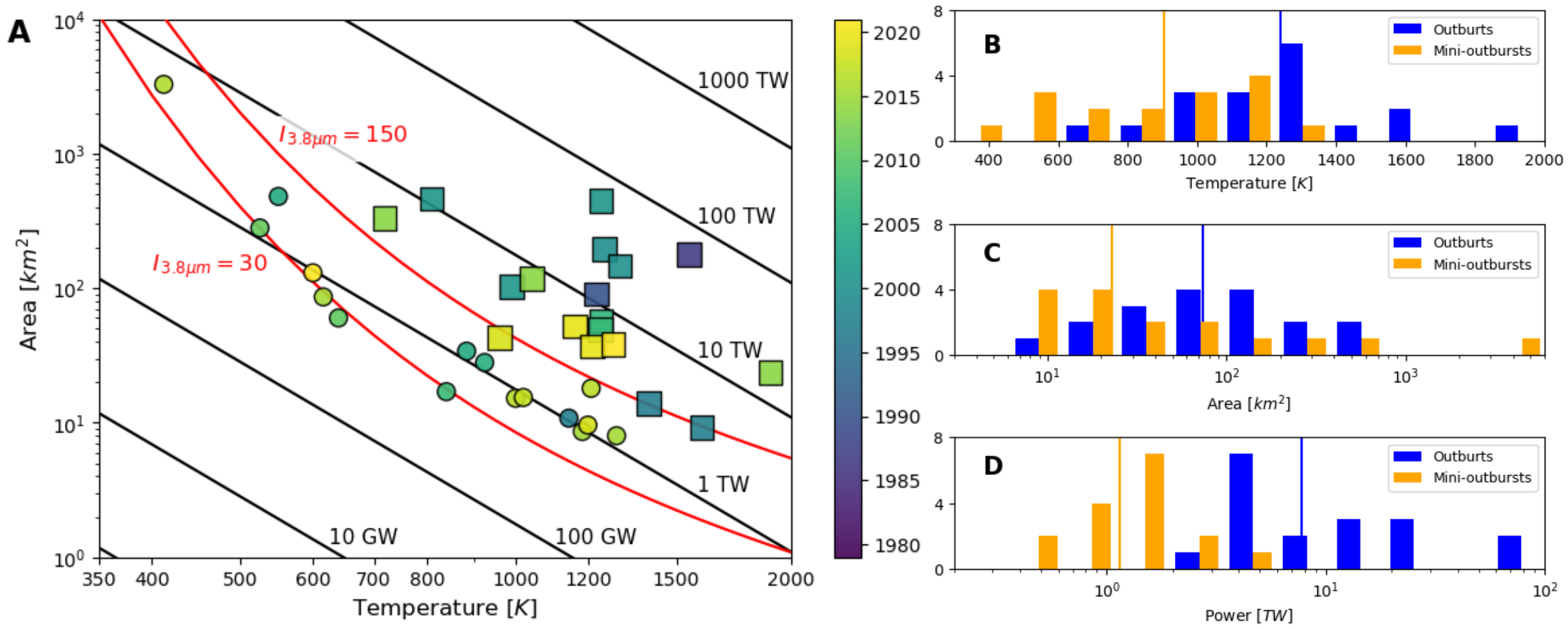

**Figure 4.** (**A**) Effective surface area versus the high-temperature fits on a log-log plot of the thermally constrained outbursts (squares) and mini-outbursts (circles) points. The black lines show the Weins-law power of the area-temperature space, and the red lines show the range of areas and temperatures that emit 30 and 150 GW/sr/µm at 3.8 µm (Lp-band). The histograms are the (**B**) effective temperatures, (**C**) effective areas, and (**D**) total powers of the eruptions. The vertical lines are the median value of each population.

## 2.2 High-level trends

After compiling a comprehensive record of Io's published outburst activity over the last five decades, several spatial and temporal patterns emerge. Importantly, many of the trends have become apparent only after a multi-decade baseline emerges to disentangle long term processes in Io's complex volcanic behavior. Many of these trends are presented here for the first time because they are difficult to discern on the sub-decadal timescales of previous studies.

### 2.2.1 Location trends

The 17 localized outbursts in this study originate from 13 unique volcanic locations. The only confirmed repeating outburst locations occurred at Tvashtar Patera and Pillan Patera for 4 and 2 localized events, respectively. There is evidence that Acala Fluctus is also a repeating outburst location. Besides these, however, the other 11 outburst locations are "one-off" outburst events. Of these unique volcanic locations, over half (or 7 of the 13) are not clearly associated with any of the emission centers observed by Galileo. The event names in Table 2 are the location names of the associated hot spots, and events without previously known hot spots have alphanumeric names assigned by the discoverer (e.g., the event names "0002A", "UP 254W" and "202010A"). By contrast, the 27 localized mini-outbursts originate from 16 unique volcanic locations, most associated with known or named hot spots. Since a disproportionate number of outbursts occur without an apparent correlation with previously observed hot spots, the volcanic mechanism responsible for the brightest transient events might not require known hot spots. Instead, a subset of outbursts could reactivate dormant volcanoes or be "one-off events" unassociated with any previous hot spot.

Figure 1 shows a relatively uniform scatter of outburst locations on the global scale. On a finer scale, however, several regions do not show outburst activity in this dataset (see Figures 1, 4, and 7). The sub-Jovian and anti-Jovian longitudes from 30W to 340W and 180W to 125W seem devoid of outburst. The South Pole has no event below 56S where the 2013 event at Heno Patera took place. The spatial distributions of outbursts and mini-outbursts could be decoupled. Except for Pillan Patera, where several of both have occurred, outbursts and mini-outbursts seem to dominate different regions.

### 2.2.2 Changing location trends

Although Figure 1 shows that outbursts have a roughly uniform global distribution over the four decades of this dataset, this pattern does not persist for shorter time periods. In fact, Figure 2 shows a stark change in outburst locations between the 1995-2012 and 2013-2022 periods which favor the leading and trailing hemispheres, respectively. Mini-outbursts, by contrast, consistently prefer the trailing hemisphere (180W to 0W) for the entire dataset. The statistical significance of these trends is explored in sections 3 and 4, and we discuss their scientific significance in section 5.

These changing trends designate three periods in the dataset. The first period between 1979 and 1994 was before many modern infrared observational capabilities. The data is sparse, and we cannot say there was much outburst behavior in that early period. What followed was a period when outbursts were primarily found in the leading hemisphere. The first was 9503A detected with IRTF (Spencer et al. 1997). This trend continued through the Galileo mission and afterward until the 2007 New Horizons flyby detected the last outburst at Tvashtar (Tsang 2014, Spencer et al., 2007). No outbursts were detected anywhere on Io between 2008 and 2012. Outbursts were observed again in 2013 but were notably in the trailing hemisphere, and that trend continued at least until late 2021.

### 2.2.3 Temporal clustering on decadal timescales

How often do outbursts occur on Io and does this frequency change over time? Table 3 summarizes the number of infrared observations and outburst detections for each of the major ground-based observation campaigns and spacecraft missions surveyed. Some small ground-based campaigns are clumped for brevity, but every effort was made to verify the total number of observations in each 1-6 year period. Table 4 further summarizes these observations into three time periods: 1997-1993, 1995-2012, and 2013-2022. The number of outbursts in Table 3 and 4 is 28 instead of the 30 recorded in Tables 1 and 2 because we excluded the unnamed January 1978 event and the late 2000 Tvashtar detection made with the Hubble Space Telescope (HST) at non-thermal wavelengths (less than 1 μm). Outbursts are rare in the sense that extensive, multi-year observation campaigns often find only a handful. While estimating the frequency from individual campaigns is prone to the uncertainties of small number statistics, the aggregate information in Table 4 is considerably more robust. Several results are surprising: first the average outburst frequency between 1978 and 2022 was 23 ± 4 per year, and second that they were significantly more frequent in the years 1995 to 2012 than before and after. We justify these claims below.

We define frequency as either the number of events in a period of time or equivalently as the probability of detecting an event in one observation that sees half of Io's surface. If $n$ stochastic events are detected in $N$ unbiased observations, then each observation has a $(n \pm \sqrt{n})/N$% chance of detecting an event and roughly twice this probability for an event happening somewhere on Io. The frequency is then about 365 * 2 $(n \pm \sqrt{n})/N$ events per year. This estimation assumes that outbursts (1) persist in a detectable fashion for about 24 hours, (2) are detectable everywhere on the visible hemisphere, and (3) are equally probable on the visible and invisible hemispheres. We use these assumptions mainly for consistency with previous studies (Spencer and Schneider 1996), and other sets of assumptions do not greatly affect the relative changes in frequency that we detect throughout the. However, if outbursts generally do not sustain $I_{Lp}$ >150 GW/sr/μm brightness for 24 hours, or if they are not equally likely on the visible and hemisphere; then these biases will skew the frequency estimate. We justify the 24-hour assumption with high cadence measurements of three outbursts in 2013 that had decay half-lives of about 0.8-1.8 days (de Kleer and de Pater, 2016a). This definition also includes several observational assumptions: that (1) events detected at the same location within 24 hours count as one outburst, (2) events at the same location more than 24 hours apart count as two outbursts, and (3) events are equally detectable anywhere on the hemisphere visible to the observer. The first two assumptions are negligible because high-cadence campaigns rarely observe Io on consecutive nights. The latter bias, however, is more serious because nearly all observation techniques are less sensitive to events on Io’s limb, especially for lower power mini-outbursts and for high-latitude events. Due to these limitations, our estimates generally underestimate the true frequency.

An early estimation of Io's outburst frequency was based on the five events found 1978 to 1995 and gave a 3.3 ± 1.5% probability or 12 ± 5 outbursts per year (Spencer and Schneider 1996). A more recent study found a 8 ± 3% frequency based on a 2016-2022 IRTF campaign (Tate et al., 2023). For the whole dataset 1978 to 2022 summarized in Table 2, the frequency is 6.2 ± 1.2% or 23 ± 4 outbursts per year. Although these values appear to be relatively persistent, outbursts are not constant on decadal or yearly timescales.

Table 4 gives frequencies of 3.4 ± 1.7%, 16 ± 4%, and 4.3 ± 1.4% for the periods 1978-1993, 1995-2012, and 2013-2022, respectively. The frequency for 1995-2012 is a factor of 3.7 higher than 2013-2022 and separated by more than a two-sigma, which we interpret as a significant change in outburst behavior. Observationally, the campaigns in 1995-2012 and 2013-2022 used similar techniques and many of the same telescopes (e.g., IRTF, Gemini N, and Keck II). We therefore cannot identify any major experimental biases that could systematically skew these results.

Mini-outbursts require more sensitivity, and we estimate their frequency from only the spacecraft flybys and ground-based AO campaigns (see Table 3). Mini-outbursts between 1996 and 2022 occurred at a higher rate of about 61 ± 11 events per year or 17 ± 3% of the time. This is 2.7 times more frequent than outbursts, and intriguingly this ratio is roughly constant as outburst frequency rises after 1995 and decreases after 2012. The mini-outburst frequency decreased from 48 ± 12% to 10 ± 3% between 1996-2011 and 2013-2022, respectively. The dataset is not sensitive to mini-outbursts before Galileo’s 1996 arrival. The campaigns that are most sensitive to mini-outburst are even less susceptible to experimental changes than the larger dataset used for outbursts. In summary, both outbursts and mini-outbursts are significantly more frequent in the years 1995/6-2012 than 2013-2022, and mini-outbursts are consistently 2-3 times more frequent than outbursts.

There are also years-long periods of this dataset when no outbursts were detected. These periods of low-outburst activity are 2002-2005, 2008-2012, and 2014-2017. The absence of detections could be because of the low observational cadence over these periods, especially the first two (Cantrall et al., 2018). However, this does not explain why the intermittent ground-based observation campaigns observing Io at these times succeeded in detecting a large number of mini-outburst events (see Figure 3). If this is not an observational bias, then mini-outburst activity seems more consistent from year to year, whereas outburst activity seems to fluctuate greatly and even cease for 3-5 year periods.

### 2.2.4 Temporal clustering on short timescales

If frequency changes on decadal timescales, how constant is it on shorter periods? While this question is more difficult to test with our dataset, there is circumstantial evidence that outbursts and mini-outbursts are also clustered in time over 1-20 day timescales. Two examples of this phenomenon are: the three rapid outbursts at Amirani, Tvashtar, and Surt observed within a span of 3 days in 2001 (Marchis et al., 2002); and the three outbursts at Heno, Rarog and 201308C, and one mini-outburst at Loki observed within 15 days in 2013 (de Kleer and de Pater, 2016a). The trio of outbursts in 2013 is especially compelling because despite frequent observations no other outbursts were detected in the five years before 2008-2012 and the fours years after 2014-2018.

One way to quantify clustering in time is to count the fraction of events that occur within a time interval of at least one other event. We arbitrarily chose an interval of 10 days. This clustering metric does not account for irregular schedules that do not always observe Io more than once in any given 10-day window. We find that 26 of 60 events 1995-2022 are clustered in time, or that 43 ± 9% of outbursts and mini-outburst events happen within 10 days of another outburst or mini-outburst. The expected clustering of independent events with randomly timed observations is no greater than 10%. Comparing this to the measured 43 ± 9%, we conclude that outbursts and mini-outbursts are significantly clustered at 10-day timescales. The results of this analysis are largely the same for the subsets of the data, including only the outbursts, mini-outbursts, both for 1995-2011, and both for 2013-2022. In each case, 30% to 50% of the events are clustered in 10-day intervals. This shows how bright transient events are more likely to occur.

### 2.2.5 Temperature trends

Outbursts are detected when relatively large quantities of high-temperature lava suddenly appear on Io's surface. Consequently, these events contain the hottest materials measured on the surface of Io (excluding the atmosphere and interior). Whether outbursts are, in fact, the hottest form of volcanism on Io or just the most visible source of high-temperature activity remains an open question. While outbursts are transiently the brightest hot spots on Io, they are not the only hot spots that reach the temperatures necessary for non-sulfurous volcanism. High-resolution measurements from Galileo's NIMS instrument show that many other hot spots reach comparable temperatures above 600K (McEwen et al., 1998; Williams and Howell 2007). Nevertheless, outbursts and mini-outbursts are significant – if not the primary – mechanisms for transporting silicate magma to the surface and resurfacing Io with non-sulfurous materials.

The effective temperatures of the outbursts in this study range from about 700K to ~1900K. Figure 4 shows the temperature-area behavior of the 33 constrained events. Some values are constrained with two-temperature fits, for which we take the hottest temperature. Mini-outbursts generally have lower effective temperatures, which could be an observational bias. Both populations have a maximum density of around 1200 K, consistent with silicate volcanism <1475 K and too hot for sulfuric volcanism <600 K (Caar, 1986; Schneider and Spencer, 2023).

Four of the 18 outburst events (22 ± 11%) have ultra-high temperatures greater than 1475K. These have sparked debate over possible ultramafic volcanism on Io (Howell et al. 1997). The first of these events was detected in 1986 at 4.8-8.7 μm (Veeder et al. 1994), a wavelength range that is not ideal for constraining high temperatures. The other ultra-high temperature events are better constrained and remain the best evidence of possible ultramafic volcanism. The 9610A outburst at ~1500 K (Stansberry et al. 1997) and the 1997 Pillan Patera outburst at ~1600 K (Davies et al., 2001; Howell et al., 2001) occurred in the late 1990s. Subsequent surveys by Galileo and ground-based AO telescopes did not detect activity temperatures >1475 K (Lopes et al. 2007). An exception to this is the 201308C event which has the highest estimated temperature at 1900K. This event has a spectrum that is difficult to interpret, and different fitting strategies give different temperatures that range between 1270K and 1900K, with the latter being preferred (de Kleer et al., 2014). Although ultramafic volcanism remains controversial in the community, the data suggest that some ultra-high temperatures do occur, but they are rare and only appear as outbursts (since no mini-outbursts or persistent hot spots reach such high temperatures). These ultra-high-temperature outbursts are likely exposing relatively large areas of hot materials.

### 2.2.6 Power and location trends

We do not find any strong correlations between outburst temperature and location. Nevertheless, location may correlate with total outburst power. There are 8 events with power greater than 10 TW, 6 of which occurred in the northern hemisphere. The significance of this trend is not yet quantified, however.

## 2.3 Repeating locations

While most outbursts and mini-outbursts occur at one-off locations, there are several notable exceptions. The only confirmed repeating outbursts took place at Tvashtar and Pillan Patera with the possibility of a third at Acala Fluctus. The repeating mini-outburst locations are at Pillan Patera, Loki Patera, and Kanehekili Fluctus.

### 2.3.1 Tvashtar Paetra

Six outbursts occurred at Tvashtar Paterae between late 1999 and May 2007. Tvashtar patera was closely monitored during and after the Galileo mission. Galileo's SSI captured the first high-resolution observation of fire fountains or lava curtains (Keszthelyi et al., 2001). After this period, ground-based AO campaigns did not detect outbursts at Tvashtar until before the New Horizons flyby in late 2007 (de Pater, 2014; de Kleer and de Pater, 2016a; de Kleer et al., 2019a). Although Tvashtar dominates the dataset in the time period 1999-2007, its location is consistent with other outbursts during the 1995-2007 interval.

### 2.3.2 Pillan Patera

The Pillan Patera area is an active site for persistent and transient emission. Pillan was classified as a persistent hot spot in the Galileo era (Lopes et al., 2007). After 2001, however, AO observations detected it less than half the time (Cantrall et al., 2018). Throughout both periods, Pillan had two outbursts and two mini-outbursts as well as several events below the 30 GW/sr/μm mini-outburst threshold (Marchis et al., 2000; Howell et al., 2001; Davies et al., 2001; de Kleer & de Pater, 2016a; de Kleer et al. 2019a; Tate et al. 2023). Pillan’s 1997 outburst was likely due to lava fire fountains (Keszthelyi et al., 2001; Davies et al., 2001; Veeder et al., 2012). Regarding the number of events, Pillan Patera is second only to Tvashtar Patera, which is notably on the opposite side of Io. Tvashtar and Pillan are at 120°W 62°S 244°W 12°S, respectively, and separated by a great circle of 120°.

In addition to Pillan's 1997 ultra-high temperature outburst, Galileo detected an abnormally low-dust “stealth plume” at that location (Geissler & McMillan, 2008). Two Prometheus-style plumes were observed at Pillan on 6/28/1997 and 11/8/1997 with Galileo SSI's limb imaging in the violet band of SSI). While plumes and outbursts are not causally linked, Pillan's ultra-high-temperature outburst could be associated with its subsequent outgassing.

### 2.3.3 Kanehekili Fluctus

Two Prometheus-style plumes were observed over Kanehekili Fluctus in 1997 and corroborated with later surface color and morphology changes (Geissler, 2008). Kankekili's plume detections suggest that it was active in the first half of the Galileo mission. Kanehekili had low activity for over a decade, with only a low-intensity Keck detection in 2001 (Marchis et al., 2002). In late 2010, Kanehekili erupted with two mini-outbursts and a sustained output of $I_{Lp}$ = 20-40 GW/sr/μm for three months (de Pater et al., 2014a; Cantrall et al., 2018; Davies et al., 2012). Keck observed Kanehekili three times between August and November 2010 with a steady temperature of T~ 520K. Since Kanehekili likely maintained this medium-temperature and high-intensity state between detections (possibly longer before and after the observations), its 2010 event is the longest-lived transient eruption. If mini-outbursts can last much longer than an average of 24 hours, then this would affect the mini-outburst frequency estimate of section 2.3.3 and likely underestimate their contribution to Io’s total heat budget.

Kanehekili was active again in late 2022 when Keck measured $I_{Lp}$ = 10 GW/sr/μm in September, and three months later, in November, JWST measured $I_{Lp}$~35 GW/sr/μm with a dominant temperature around 600K (de Pater et al., 2023). JWST measured the 1.7-5.3 μm spectral intensity and found a 1.707-μm SO enhancement centered on Kanehekili, evidence for plume activity. The combined evidence of plume and volcanic activity in 1997 and 2022 with the medium-temperature mini-outbursts in 2010 and 2022 suggests that each eruption was remarkably similar. If so, then Kanehekili Fluctus could have a ~12-year cycle, with the next mini-outburst and plume activity expected in the mid-2030s.

### 2.3.4 Loki Patera

Loki is the most active hot spot on Io and responsible for 10% of Io's total heat output, about 20% of the combined thermal output of all volcanic processes on Io (Veeder et al., 2012). The decades of monitoring Loki Patera illustrate the value of long-term observation campaigns and the surprising insights that such a dataset makes possible (Rathbun et al., 2002, 2006; de Pater et al., 2016b; de Kleer et al., 2017; Bartolić et al., 2022; de Kleer & Rathbun, 2023). Loki's highly periodic emission reaches the mini-outburst threshold of 30 < $I_{3.8μms}$ =< 150 GW/sr/μm every 400-550 days. This transient and predictable activity at Loki Patera is often treated as a special case, distinct from all other hot spots (Rathbun et al., 2002, Rathbun and Spencer, 2006; Davies et al. 2015). However, a similarly semi-regular activity could be more common in other hot spots. As the temporal baseline grows and probes longer timescales, we may find that Loki is merely the most energetic example of a population of periodic hot spots, which could be due to the large size of the patera. Even though Loki is the most frequent mini-outburst hot spot in our list, including or excluding it does not change the significant temporal and spatial patterns we observe. For simplicity, we include Loki observations of $I_{3.8μm}$ > 40 GW/sr/μm (primarily from the de Pater et al., 2017 dataset).

We note that there has never been a confirmed outburst observed at Loki. Despite the outburst detected in 1990 (Veeder et al., 1994), this event was not precisely localized in latitude. This event was likely not Loki Patera because it has not had an outburst in the following thirty years of careful observation (Veeder et al., 2012; de Kleer & Rathbun, 2023). Given the subsequent detections of several outbursts at longitudes near Loki (such as Acala Fluctus), the 1990 outburst likely originated from some other location.

### 2.3.5 Acala Fluctus Area

Acala Fluctus is a large volcanic region west of Loki Patera and the closed outburst location to thereto. Acala showed little thermal activity before its two 2019 outbursts (Tate et al., 2023). These took place in May and June 2019, a month and a half apart, with lower intensity (~2 GW/sr/μm) detections before, between, and after the eruptions.

# 3 Statistical Tests

We establish statistical tests to ascertain whether large transient events are randomly distributed on the surface of Io. Prior investigations have mainly employed two tests: mean absolute latitude and mean pairwise spacing, to accept or reject a null hypothesis stating that the population is randomly distributed on a sphere (de Kleer and de Pater 2016b; Cantrall et al., 2018). In addition to applying these heritage techniques, we also investigate the spatial distribution of outbursts and mini outbursts relative to the axes of a three-dimensional cartesian coordinate system where the X-axis goes through the anti / sub-Jovian points and the Z-axis goes through Io's poles. This coordinate system is convenient to assess hemispherical dichotomies (e.g., leading/trailing hemisphere) and quantify the confidence that any observed trends are real and not the result of a random distribution. We employ Kolmogorov–Smirnov (K-S) tests to compare the observed outburst and mini-outburst populations to each other, as well as to simulated populations from defined probability distributions (e.g., random). Compared to previous studies, this analysis uses a larger dataset that spans a greater range of time.

Note that we did not apply the nearest-neighbor distance analysis used by Hamilton et al. (2013) and Tyler et al. (2015), as that study analyzed hundreds of hot spots and patera locations and would not be appropriate on the smaller sample size of our dataset (with *n* between 5 and 44) .

## 3.1 Mean Absolute Latitude

The mean absolute latitude $\underline{|\varphi|}$ tests if a set of points has a polar or equatorial preference compared to a random collection. To know if $\underline{|\varphi|}$ is consistent with a random population, it is compared with a collection of *n* randomly selected points. Random points have a mean absolute latitude of $\underline{|\Phi|} \simeq 32.7 \pm 32.7/\sqrt{n}\,^{\circ}$ degrees, with this uncertainty valid only for large *n*. We compute the n-dependent uncertainty value for each set directly from the standard deviation of $N$=10,000 simulations of *n* random points on a sphere. The result of this statistical test is consistent with the null hypothesis if it has percentiles between >5%ile and <95%ile. The test provides evidence for a polar preference if >95%ile and the opposite for <5%ile.

For example, to quantify the likelihood that the outburst and mini-outburst events (n=44) have a polar preference, we can compute the actual or observed mean absolute latitude $\underline{|\varphi|}$ = 31.6 and a distribution of expected $\underline{|\Phi|}$ = 32.7 ± 5.4° for $N$=10,000 random sets of n=44 points each. Figure 5 shows the distribution of random simulations, which is nearly Gaussian for large *n* values. The actual value for $\underline{|\varphi|}$ is in the 38th percentile in the $\underline{|\Phi|}$ distribution of simulations. The mean absolute latitude statistic for these values is consistent with random locations. Therefore, the n=44 outbursts and mini-outbursts do not have a significant polar preference. We repeat this analysis for each subset of *n* points specified in the columns of Table 5 and record the percentiles in the row labeled “mean absolute latitude.” We then repeat this for the other statistical tests that are listed in the rows of Table 5 and explained in sections 3.2 and 3.3. We call these “percentile of the mean " tests. Because the percentile is size-independent for various values of *n*, it is more informative than the calculated mean value. For this reason, Table 5 only reports the percentile values so that the statistical results of various sample sizes can be more easily juxtaposed.

## 3.2 Comparison to Io-centered fixed body coordinate system

To generalize the mean absolute latitude test for the other directions, we analyze points on the globe of Io in cartesian coordinates (see Figure 3). The X, Y, and Z axes extend from anti-Jovian (0 W) to sub-Jovian points (180 W), leading (90 W) to trailing (270 W), and from the South pole to the North pole, respectively. We also compute the mean absolute values long these axes, |x|, |y|, and |z|, to measure the symmetry between the 90th meridian, prime-meridian, and equator planes, respectively. For a set of *n*

longitude and latitude coordinates on Io, we project each point into $\mathbf{x}_i = (x_i, y_i, z_i)$ coordinates (Figure 2) before calculating the actual mean and standard deviation values for each direction. As explained above, we compare this measured mean to the averages of a large number of randomly generated simulations $\underline{X}$ and calculate the percentile of the measured mean $\bar{x}$ relative to those simulations. The expected mean of the values $\underline{X}$, $\underline{Y}$, and $\underline{Z}$ are 0.0 $R_{Io}$ or 0.0° from the plane of symmetry (equator, prime-meridian, or 90th meridian). Conversely, the expected mean of the *absolute* values $\underline{|X|}$, $\underline{|Y|}$, and $\underline{|Z|}$ are 0.5 $R_{Io}$ or 30.0° from the plane of symmetry.

To continue the above example for mean absolute latitude, we can similarly quantify the polar preference of the n=44 outbursts and mini-outbursts by taking the mean absolute projection along these axes. This value is $\underline{|z|}$ = 0.489 $R_{Io}$ which we compare to the distribution of random simulations $\underline{|Z|}$ = 0.500 ± 0.064 $R_{Io}$. This puts $\underline{|z|}$ in the 40th percentile of the simulated |Z| values, meaning that the n=44 events do not have a significant polar preference. Since the $\underline{|\varphi|}$ and $\underline{|z|}$ tests both measure polar preference, the percentile values are approximately equal $\underline{|z|} \gtrsim \underline{|\varphi|}$. Note that while the mean value lange the Z-axis is $\underline{|Z|}$= 30° latitude, this is slightly smaller than the $\underline{|\Phi|}$ = 32.7° latitude for the mean absolute latitude (compare the percentile values for $\underline{|z|}$ and $\underline{|\varphi|}$ in Table 5).

## 3.3 Mean Pairwise Spacing

Another test for randomness is the mean pairwise spacing, $\underline{d}$. Instead of projecting each point along a global axis, this method calculates the great-circle distance between every pair of points; d( $\mathbf{x}_i, \mathbf{x}_{i'}$) for $i \neq i'$. As above, this mean value is compared to the distribution of simulated values. This test measures any clustering or repelling nature of the points. As above, a set of points with a mean $\underline{d}$ value that is far from the mean (50%ile). We consider the set clustered for small percentiles (<5%ile) and repelled for large percentiles (>95%ile).

## 3.4 Kolmogorov–Smirnov Tests

We also use the Kolmogorov–Smirnov (K-S) test (Massey, 1951; Virtanen et al., 2020) to determine the goodness of fit or confidence that the set of points is consistent with a set drawn from a random population. We also compare two sets of points to determine the confidence that they come from the same population (Hodges, 1958; Virtanen et al., 2020). This statistical test is similar but somewhat more robust than the percentile of the mean statistics used above. We present both results because they measure different properties of the dataset. Since the K-S test is one-dimensional, we calculate the confidence in each XYZ direction, from which we take the largest value to be the lower limit of the combined confidence in all three dimensions.

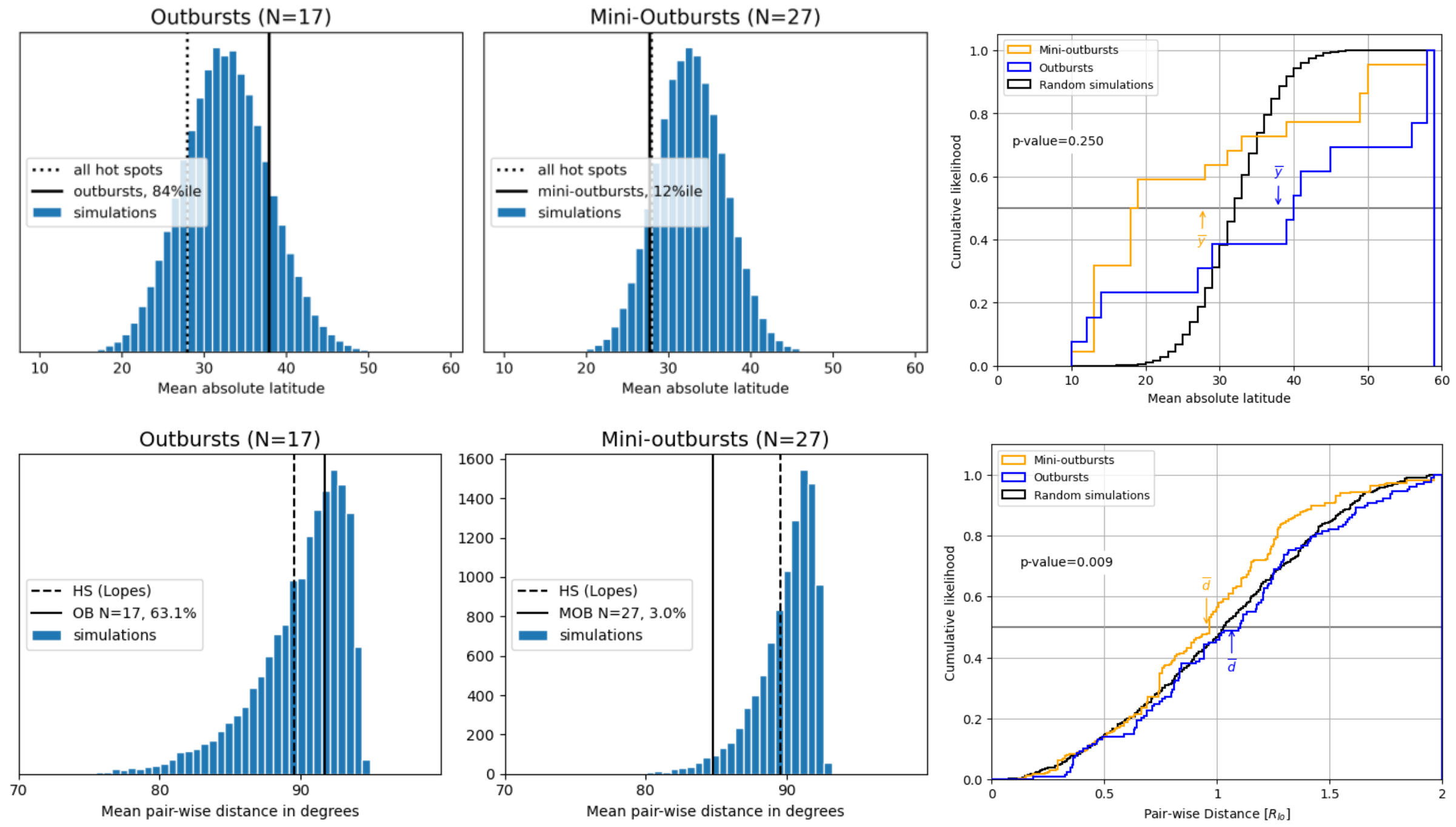

**Figure 5.** Io's spatial distributions of mini-outbursts (left) and outbursts (middle). The mean absolute latitude (top) and mean pairwise distance (bottom) show the statistical behavior of each relative to the histograms of ten thousand random simulations.

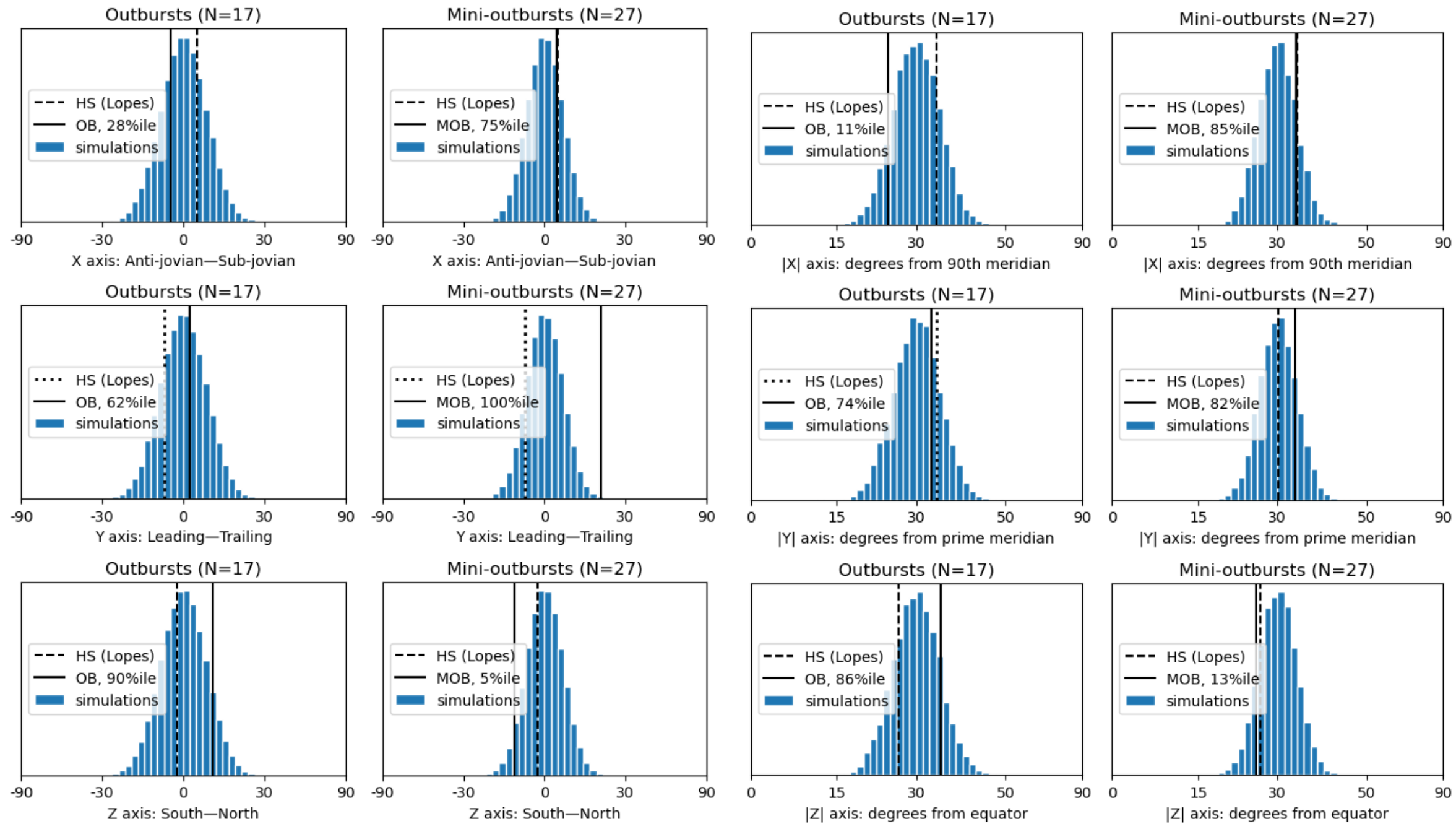

**Figure 6.** Statistical tests to determine the probability that the observed distribution of outbursts and mini-outbursts are random. The blue histogram in each subplot represents results from 10,000 random simulations, and the solid black vertical line is the mean value of the outbursts (bottom) or mini-outbursts (top). The dashed black lines are the averages for all known hot spots.

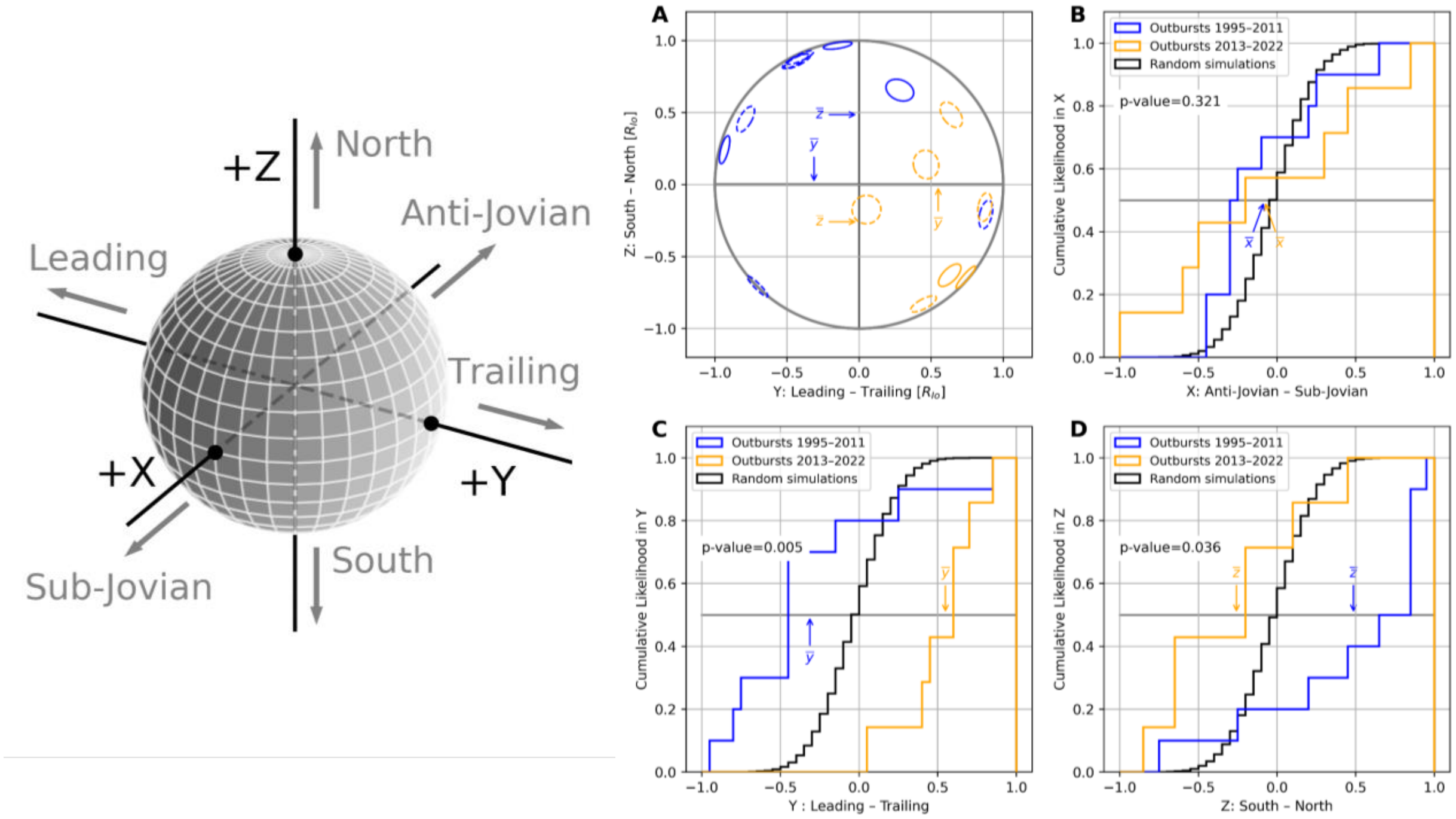

**Figure 7**: The outbursts 1990-2011 (red) and 2013-2022 (blue) are different populations. **A** shows the outburst locations projected on the Y-Z plane (north is up, south down, leading left, trailing right). The ovals are solid when the outburst is in the sub-Jovian hemisphere and dashed for the anti-Jovian hemisphere. Subplots **B**, **C**, and **D** show the cumulative histograms in the X, Y, and Z directions. The K-S p-value confirms that the outbursts before and after 2012 are samples from the same population.

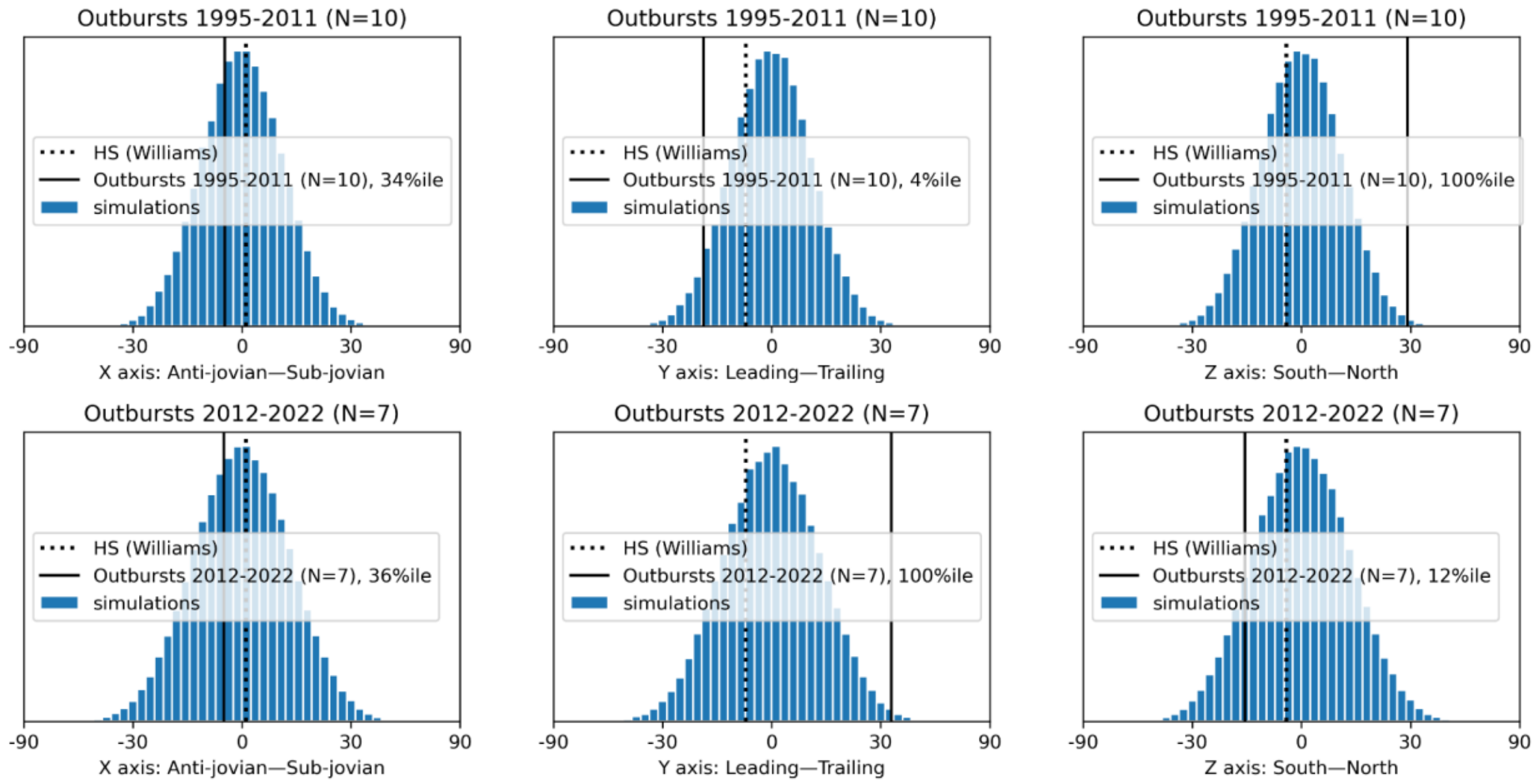

**Figure 8.** Plot of outburst statistics for years 1995-2011 (top) and 2012-2022 (bottom).

| | Outbursts | | | Mini-outbursts | | | Both | | | Hot spots |
|---|---|---|---|---|---|---|---|---|---|---|
| Time range | **1995-2011** | **2013-2022** | **all** | **1995-2011** | **2013-2022** | **all** | **1995-2011** | **2013-2022** | **all** | |
| Sample *n* | 10 | 7 | 17 | 14 | 13 | 27 | 24 | 20 | 44 | 275 |
| Hemi pref X | 34% | 36% | 28% | **96%** | 20% | 75% | 85% | 18% | 57% | 4.8° |
| Hemi pref Y | **4.3%** | **99.7** | 62% | **99.9%** | 91% | **99.9%** | 89% | **99.4%** | **99.6%** | -2.2° |
| Hemi pref Z | **99.7%** | 12% | 90% | 14% | 11% | **4.8%** | 82% | **4%** | 30% | -2.3° |
| Hemi sym \|X\| | 2% | 70% | 11% | 89% | 58% | 85% | 36% | 68% | 51% | 34.1° |
| Hemi sym \|Y\| | 69% | 68% | 74% | 95% | 38% | 82% | 94% | 51% | 88% | 30.1° |
| Hemi sym \|Z\| | **97%** | 28% | 86% | **0.4%** | 86% | 13% | 21% | 70% | 40% | 26.6° |
| Mean abs lat \|$\phi$\| | 96% | 26% | 84% | **0.3%** | 83% | 12% | 22% | 65% | 38% | 28.5° |
| Mean pair dist | **1.0%** | **4.0%** | 63% | **1.0%** | 28% | **2.6%** | 52% | 1.6% | 20% | 89.6° |
| Mean pair dist* | 6.3% | **4.0%** | **99.9%** | **2.0%** | 35% | **4.3%** | **99.9%** | **1.9%** | 53% | 89.6° |
| Summary | *leading, northern, poleward pref., clustered* | *trailing hemi. pref., clustered* | *no pref. or sym., repelling** | *trailing and equatorward pref., clustered* | *no hemi. pref., random* | *trailing southern hemi. pref., clustered* | *no pref. or sym., repelling** | *trailing, clustered* | *trailing, random* | *leading, random* |

**Table 5.** Statistical results for the spatial tests. These percentiles are where the mean of each test mean value is in a distribution of random simulations of the same sample size.

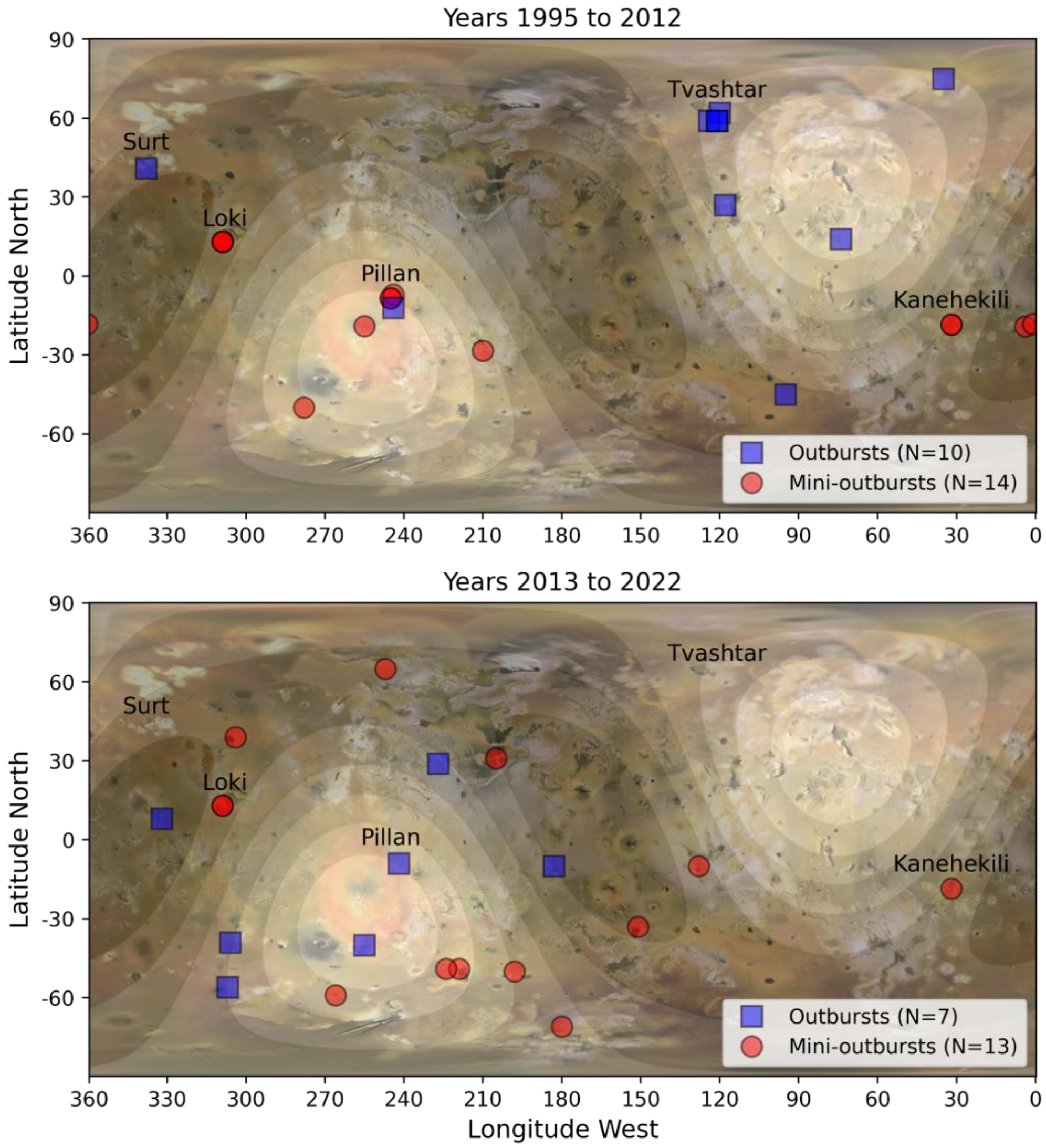

**Figure 9.** Same as Figure 2 plotted with the density of mountains from the l=2 spherical harmonics coefficients from Kirchoff et al. (2011). The light areas have significantly more mountains than the dark areas. Outburst locations (blue) and mini-outburst locations (red) for 1995-2011 (top) and 2013-2022 (bottom).

# 4 Interpretations and Results

Table 5 gives the percentile scores for spatial statistical tests described above. These tests are evaluated for three time intervals. One is the whole dataset of localized events 1995-2022, and the others divide the dataset before and after the year 2012. The choice for these dates is discussed in section 4.3.

## 4.1 Global trends over the whole dataset

The most statically robust result of these tests is the mini-outburst 1996-2022 preference for the trailing hemisphere ($\bar{y}$ =99.9%ile, n=27) (see Figures 1, 4 and 6, and Table 5). Secondarily, mini-outbursts have southern preference ($\bar{z}$ =4.8%ile, n=27). Outbursts on the other hand do not show a leading-trailing ($\bar{y}$ =62%ile, n=17) or a significant north-south preference ($\bar{z}$ =90%ile, n=17). Section 4.3 explores the significant temporal change of outburst preference along the leading-trailing axis.

## 4.2 Polar trends and the lack thereof

Previous studies found that the outbursts and mini-outbursts of 2013-2018 have spatial distributions that are clustered at high absolute latitudes (de Kleer and de Pater, 2016b; Carntrall et al., 2018). While we confirm these results for the specific AO observations of 2013-2018, this trend is not persistent over longer time periods, but instead more generally consistent with random locations. Considering the entire dataset over multiple decades, outbursts and mini-outbursts do not significantly occur at high absolute latitudes (or closer to the poles). Nevertheless, it is interesting to note the weak but opposite polar behavior for outbursts ($\underline{|z|}$ = 86%ile, n=17) and the equatorial behavior of mini-outbursts ($\underline{|z|}$ = 13%ile, n=27).

Although we do not find a poleward trend in the total n=44 dataset, this analysis treats each event the same without accounting for each event's peak temperature and power. Previous studies hypothesized that events near the poles could be more violent with higher temperatures because they are driven by deep-mantle heating and the opposite for volcanos near the equator (Hamilton et al., 2013; Tyler et al., 2015; de Kleer & de Pater, 2016b). The available data do not significantly confirm this hypothesis because many powerful, high-temperature outbursts occurred near the equator. Although low-temperature events are generally rare at high absolute latitudes, we do not find any significant correlation between temperature and latitude. A tentative link between outbursts and deep mantle heating is that outbursts have high $|\bar{z}|$ (86%ile, n=17) and mini-outbursts have low $\underline{|z|}$ values (13%ile, n=27). Although these percentiles are statistically consistent with random, the K-S test gives a 96% confidence that these are different populations.

## 4.3 Temporal trends and periods of distinct behavior

The spatiotemporal patterns of bright transient eruptions seem to change between three time ranges: 1978–1994, 1995–2012, and 2013–2022. We propose these periods because they appear coincident with changes (see Figure 3), even though their precise start and end dates are limited by the intermittent observational campaigns.

**1978–1994** (16 years)

- Outbursts (n=5) had *no clear hemispherical preference*. None were detected in the years 1980-1985 and 1991-1994.
- Mini-outbursts were not characterized due to the low sensitivity.

**1995–2012** (18 years)

- Outbursts (n=15) were more frequent and preferred the *northern and leading hemispheres*. None 2001-2005 and 2008-2012.
- Mini-outbursts (n=17) were more frequent and preferred the *trailing hemisphere and equatorial region*.

**2013–2022** (10 years)

- Outbursts (n=10) preferred the *trailing hemisphere* (and weakly the southern). None 2014-2017.

- Mini-outbursts (n=19) weakly preferred the *trailing hemisphere*.

The outburst behavior underwent a significant transition between 1995-2007 and 2013-2021. During the first period, outbursts predominantly took place in the northern and leading hemispheres of Io, while during the second period, they favored the trailing southern hemisphere. This transition happened sometime between 2006-2013, as constrained by the 2006-2007 outbursts at Tvashtar Patera and the 2013 outbursts at Heno Patera and Rarog Patera. However, the exact time of the transition cannot be specified more precisely due to a multi-year absence in outburst activity from 2008-2012. Interestingly, mini-outbursts during this time continued at approximately the same frequency and were more likely to repeat from locations near the equator.

A spatiotemporal trend can be observed in the outburst behavior. Outbursts before 2012 favored the leading hemisphere while those observed after 2012 favored the trailing hemisphere. The outbursts of 1995-2007 were found on the northern and leading hemispheres while those of 2013-2022 were found on the southern and trailing hemispheres. The locations of outbursts and mini-outbursts during these two periods appear to be strongly decoupled.

The Kolmogorov–Smirnov (K-S) test was used to rigorously demonstrate that outbursts for at least two distinct populations separated around the year 2012 with a 99.5% confidence to reject the null hypothesis stating that outbursts before and after 2012 are drawn from the same population. The distribution of outbursts before and after 2012 are most different in the leading-trailing axis and most similar towards and away from Jupiter.

In addition to the outburst behavior on Io's surface, the behavior of mini-outbursts was also analyzed. The Kolmogorov–Smirnov (K-S) test was used to determine if mini-outbursts changed spatial patterns over the same time periods as outbursts. The maximum confidence of the mini-outburst K-S statistics was 86%, which is insufficient to reject the null hypothesis. Therefore, it was found that mini-outbursts do not change spatial patterns over the same time periods as outbursts.

The difference in behavior between outbursts and mini-outbursts could be due to rheological differences in the upwelling magma or due to thickness and strength differences in the overlying lithosphere through which the magma must penetrate. This suggests that there could be a barrier preventing the less powerful mini-outbursts from breaking the surface.

# 5 Discussion

## 5.1 How homogeneous are transient eruptions in time and space?

Over what timescales are the global patterns of outbursts and mini-outbursts statistically unchanging? Given the significant variations of outburst behavior throughout the historical record from 1978 to 2022, the lower limit of this timescale is about 30 years. At the high end, this could be much longer than the 45-year observational baseline if this dataset does not capture Io's primary rhythms. The 30-year low limit is estimated from the 15 years between 1995 and 2012, during which outburst spatial and frequency behavior significantly differs from the later 10 years, 2013 to 2022.

Both mini-outbursts and outbursts were about 4 times more frequent in the years 1995 to 2011 than in the decades before and after. Taken in light of the changing outburst location trends, the 1995-2011 outbursts preferred the leading hemisphere and were 3-4 times more frequent than the 2013-2021 outbursts that preferred the trailing hemisphere. Although mini-outbursts were similarly more frequent during the same period, they did not switch hemispheres. This could imply that the cause of the greater frequency in 1995-2011 is unrelated to the cause of the changing outburst locations.

Alternatively, instead of the dichotomy between the frequently outbursting leading hemisphere in 1995-2011 and the moderately outbursting trailing hemisphere in 2013-2021, it is equally valid to interpret the data as semi-constant activity in the trailing hemisphere for both periods 1995-2021 and a highly active episode in the leading hemisphere 1995-2011. Support for this alternative interpretation is that 2 of the 10 outbursts 1995-2011 are in the trailing hemisphere (Pillan 1997 and Surt 2001) counter to the overall outburst trend at that time. The top map of Figure 2 shows these locations. With the 1995-2011 outbursts being 3.7 times more active than 2013-2021, the two outbursts at Pillan and Surt could represent a similar rate of outbursts in the trailing hemisphere occurring throughout the dataset.

An early observation campaign in 1979-1981 noticed that the trailing hemisphere was generally more active at thermal wavelengths and contained six of eight possible outburst detections (Sinton et al., 1983). This is consistent with the May 5 and July 9, 1979, detections that were observed by Voyagers 1 and 2. Although the Voyager instrument suite was not sensitive to outbursts, these flybys provided visual evidence of high activity in the trailing hemisphere from longitudes 320 W to about 110 W (McEwen & Soderblom, 1983). If this activity was due to bright transient eruptions, this longitudinal distribution was likely more similar to the 2013-2022 era than the intermediate Galileo-era.

Since the 1995-2011 period differed from what came before and after, Io's bright transient eruptions could alternate between the leading and trailing hemispheres on a ~30-year timescale. Although the evidence for a global oscillation of Io's outburst behavior is presently weak, this hypothesis predicts that outbursts will transition to the leading hemisphere sometime in the 2030s. Even if something like this cycle is occurring on Io, however, it is difficult to form a physical hypothesis to explain this behavior (see Section 5.2).

Despite the uncertainty of how these findings constrain Io's geology, they have strong implications for interpreting short-duration observation campaigns or spacecraft missions. For instance, the detailed portrait of Io's volcanism captured by the Galileo mission from 1996-2001 might not represent Io's long-term behavior. This is especially true for >50-year variations. Several persistent hot spots observed during the Galileo era have disappeared and new ones have appeared elsewhere (de Kleer & Rathbun, 2023), suggesting that the evolution of persistent hot spots is slow but observable. Although hot spots fluctuate, their global distribution appears constant (de Kleer et al., 2019a). What makes outbursts unique is their variability on every timescale that we can measure.

## 5.2 What could cause the leading-trailing change?

Although outbursts appear homogeneous (or repelling) distributed over Io's surface during the 1978-2022 observational baseline, they switch from a leading hemisphere preference in 1995-2011 to trailing in 2013-2022. This is in contrast with mini-outbursts, which always prefer the trailing hemisphere. From a theoretical perspective, this dichotomy between the leading and trailing hemispheres is unexpected. Two prevalent surface heat flux models – derived from the deep-mantle and asthenospheric heating models – do not have leading-trailing asymmetries to explain this (Hamilton et al., 2013; Tyler et al. 2015). The only surface heat flux asymmetries in these models come from second-order effects along the sub-Jovian to anti-Jovian axis. However, since neither Io's hot spots nor dormant paterae follow these ideal heat flux models (Hamilton et al., 2013; Tyler et al., 2015), it is of little surprise that large transient eruptions are also different.

One of Io's asymmetries between the leading and trailing hemispheres involves the Io plasma torus. A 53–57 km/s tailwind bombards Io's trailing hemisphere and causes a slight atmospheric density enhancement in the leading hemisphere (Walker et al., 2010; Blöcker et al. 2018; Bagenal & Dols, 2020). Although Io's atmosphere is complexly nonuniform with larger asymmetries between day and night, sub-Jovian and anti-Jovian, and pole and equator, the plasma headwind could differentiate the geology of these hemispheres (de Pater et al., 2021, 2023). Perhaps this affects the nature of volatile deposition on Io's surface. However, the SO2 ice distribution does not show a strong leading-trailing asymmetry (Trumbo et al., 2022; de Pater et al. 2023).

Another hypothesis is that outburst activity is related to Io's mountain-building regions. This idea was briefly proposed by Cantrall et al. (2018) to describe the trailing bias of the 2001-2016 bright transient events. The present discoveries make this theory more plausible. The bimodal (k=2) clustering of Io's mountains is remarkably similar to the outburst spatiotemporal patterns described above. Kirchoff et al. (2011) used spherical harmonic fitting to analyze the global pattern of mountain locations. They discovered that the mountains are highly concentrated around two main locations. One on the leading hemisphere at 20N 80W and one on the trailing hemisphere at 15S 260W, and that these locations are anti-correlated with hot spot locations (see Fig. 3 of Kirchoff et al. (2011) and Fig. 4.10 of Keane (2023)). These findings are confirmed by further analysis of Io's mountain distributions (White et al. 2014; Ahern et al. 2017; Keane et al., 2023). Importantly, the global clusters of mountains are near the mean outburst locations during the periods 1995-2011 and 2013-2022, respectively. The cluster around 20N 80W is denser, which could explain why the 1995-2011 outbursts were 4 times more frequent.

The high tectonic stresses in the mountainous regions can cause powerful seismic shocks that might trigger outburst eruptions. Galileo found that Io's mountains are surprisingly tall, so much so that they set stringent constraints on Io's lithospheric composition, temperature, and thickness (Turtle et al. 2007; White et al. 2014; Bland and McKinnon, 2016; Keszthelyi et al. 2023 and references therein). We hypothesize that slowly ascending magma builds pressure and energy in the upper lithosphere until a nearby seismic event ruptures the magma chamber and provides a path for its rapid ascent, although we recognize that future modeling efforts will be required to substantiate this possibility.

The conditions needed for an outburst include the formation of high-pressure magma near the surface and the trigger that quickly releases this magmatic energy by transporting large volumes of high-temperature lava to Io's surface. We can call these two conditions “outburst potential” and “outburst trigger”, respectively. The triggering mechanism could be io-quakes or landslides generated in Io's mountain-forming regions. After these conditions are met, there is an “outburst signature”, namely the sudden and short-lived infrared enhancement produced by the high-temperature lava. In this hypothesis, therefore, mountains are somehow critical for creating the conditions necessary for outbursts.

Judging from the distribution of outbursts, the pockets of high outburst potential would accumulate over decades in a fairly uniform pattern over the globe of Io. Both the outburst potential and the triggering energy would need to reach a critical threshold before an outburst takes place. The sizes of these pockets would vary to account for the two orders of magnitude range of power between mini-outbursts and the largest outbursts. The magma scavenging effect (Hamilton et al. 2013) and the time required to produce these pockets would repel each other. Once outburst potential accumulates, however, a common trigger or series of triggers would act on all nearby pockets. Exploring this parameter space would be a valuable line of future research. The apparent scarcity of powerful triggers would counteract the magma-scavenging effect over long distances. This balance of two opposing effects could also explain why bright transient events are both temporally clustered in ~10-day intervals and spatially repelled over ~20-year intervals.

Assuming that mountains are associated with outbursts, we can infer that the northern leaning cluster of mountains near 20N 80W was highly active in the years 1995-2011 followed by a dormant period from 2013-2022. At this time, the mountains in the trailing southern hemisphere centered at 15S 260W became more active. This would imply that seismic triggers act over large distances ~2000 km. To explain mini-outbursts with the same mechanism, the more spread-out cluster of mountains in the trailing hemisphere must be more amenable to lower energy events than the dense cluster of mountains in the leading hemisphere. Terrestrial mountains are associated with both the dampening of small earthquakes and the amplification of powerful earthquakes (Meunier et al., 2008; van der Elst et al., 2016; Weber et a l., 2022; Li et al. 2019), and Io's mountains in the leading hemisphere may preferentially make the conditions for only large outbursts. The dense mountains would require higher outburst potential before the triggering threshold is met. If true, this means that outburst activity provides an indirect measure of Io's seismic activity, which appears to alternate between leading and trailing hemispheres on ~30-year timescales.

## 5.3 How are outbursts associated with persistent hot spots?

Like outbursts, persistent hot spots also changed behavior between the Galileo-era (1996-2001) and AO-era (2001-2016) (Cantrall et al. 2018). Unlike outbursts, however, this transition seemed to start earlier and last longer. If mountain-forming is not causally related to outbursts, then perhaps correlated changes in other volcanic activity can inform what a common mechanism would look like.

Cantrall et al. (2018) defined a persistent hot spot as something detected in more than half of the observations capable of detecting it. They found 18 such locations: 5 on the leading and 13 on the trailing hemisphere. An examination of the Galileo detections shows approximately 13 locations that meet this criterion: 5 in the leading and 8 in the trailing (Lopes et al. 2007). This is a net gain of five (or a 43 pm 17% increase in) persistent hot spots in the trailing hemisphere over the course of ~10 years. Although the transitional time frame varies for each hot spot, the first changes were visible around 2000 near the end of the Galileo era and the latest emerging hot spots established their full brightness around the year 2013 (see Fig 12. of Cantrall et al. 2018).

Loki Patera is the most extravagant example of this transition. Loki's time series (see Fig 6.10 of de Kleer and Rathbun, 2023) shows a significant difference between a 540-day cycle before 2002 and a ~480-day cycle after 2013. Between 2002 and 2009, Loki was in a low-brightness transitional state, followed by erratic episodes until 2013 when it returned to a periodic output even brighter than before. Loki follows the trend of the persistent hot spot population by dimming after 2000 and ramping back up to higher and more predictable levels in the early 2010s.

The increase of persistent hot spots in the trailing hemisphere could be related to a similar switch in the outburst hemispheres. For 2000-2013, the combined effect of fewer persistent hot spots in the trailing hemisphere and four times as many outbursts in the leading hemisphere might have a common cause. To compound this trend, Loki emitted significantly less energy from the trailing hemisphere in the 2003-2009 timeframe. This asymmetry in Io's heat flux might be measurable if we took a closer look at the entire dataset. Although inferences of Io's total power do not show a significant heat flux difference between the leading and trailing hemispheres, these are indirect measurements based on the number of discernable hot spots and patera distributed over Io (Veeder et al. 2012; Cantrall et al. 2018). A more detailed, time-dependent heat budget solution would likely reveal global changes in total volcanic radiation that coincide with changes in outburst patterns. If true, this could lead to a common mechanism for decadal variations in Io's volcanism irrespective of volcanic style.

## 5.4 Future analysis

There are several aspects of this dataset that remain to be explored. A rigorous correlation between transient and persistent hot spot locations would be of great value, as would a comparison between transient hot spots and mountains (analogous to Kirchoff et al., 2011). We did not explore the 2-5 µm ratio of thermal emission, which could expand on previous discoveries of how outbursts differ from other volcanic styles (Davies et al. 2010). Future investigations can also expand the time clustering analysis in three ways: first, by factoring in the specific observation dates and detection dates with their respective observer geometries for each campaign to control for irregular sampling rates and hemispherical preferences; second by evaluating the significance of time clustering in a broad, semi-continuous range of timescales between decades and days; and third by correlating this temporal clustering with nearest-neighbor clustering to evaluate if bright transient events trigger each other or are responding to an independent mechanism.

As this dataset grows and data science advances, future analysis will require more advanced techniques. The application of machine learning (ML) will become more necessary. ML's pattern-finding ability will likely reveal surprising correlations in Io's behavior. Whether or not these new methods transform our knowledge of Io, both these and conventional methods would greatly benefit from larger datasets with a longer observational baseline. Therefore, continued monitoring of Io is crucial.

## 5.5 The importance of more data

The most critical future work is to maintain and improve a regular cadence of high-quality observations of Io. Because Io's volcanism is highly variable and cannot yet be explained theoretically, more data is paramount to future discoveries. Our survey of the past 45 years emphasizes the critical importance of continuous, state-of-the-art monitoring of Io's volcanic activity to achieve a more comprehensive understanding. We demonstrate that volcanic activity varies significantly across decades, and the underlying processes are likely far more complex than a square pattern with a ~30-year period.

Io's volcanic timescales are several orders of magnitude faster than terrestrial planets. This attribute – namely regular volcanic events – likely holds a wealth of insights for volcanism in general. Io's surface exhibits a level of activity unparalleled by any other object in our solar system. The frequency of Io's volcanic events is comparable to the large atmospheric events on other planets (e.g. Earth's large tropical storms), emphasizing the unique opportunity to study geological processes at an accelerated pace. If Io serves as a geological analog for terrestrial volcanism, then a millennium of Earth's volcanic activity can be witnessed in a single year on Io.

But it is important to remember that Io's geological timescales are long relative to that of spacecraft missions, observational campaigns, or even human lifespans. Despite the accelerated pace of Io's volcanism, decades of high-resolution observations are required to discern the primary cycle of this dynamic moon. Io changes in a wide range of timescales: from the decadal (or 5000-day) global trends of outbursts, to the ~500-day periodicity of Loki Patera's brightness, to the 10-day temporal clustering in large transient eruptions. The complex interplay of these various timescales underscores the necessity of long-term, high-cadence, and high-quality monitoring campaigns.

# 6 Conclusions

This work comprehensively explores the behavior of Io's directly detected volcanic outbursts. The dataset we compiled shows that the localized outbursts appear uniformly distributed on Io's surface. However, this uniformity does not hold for temporal subsets of the dataset. We identified a significant change in the outburst locations before and after 2012. Random spatiotemporal distributions do not accurately characterize Io's powerful transient volcanic eruptions. Instead, outbursts cluster in specific regions at certain times. This change is a clue to the underlying processes causing Io's outbursts, which might be correlated with dense clusters of mountains.

Mini-outbursts by contrast have a more constant spatial distribution that differs from outbursts – mini-outbursts are significantly clustered in the trailing hemisphere where mountains are less frequent. More data and analysis are necessary to understand whether different mechanisms are responsible for mini-outbursts. Locations like Pillan Patera show that outbursts, mini-outbursts, and semi-persistent hot spots can occur at the same location, but this is rare. Most outbursts and mini-outbursts belong to separate locations.

Furthermore, these findings highlight the importance of continuous, state-of-the-art monitoring of Io's volcanic activity. Io's volcanism is highly variable and cannot be explained theoretically, and more data is paramount to future discoveries. Our survey of the past 45 years emphasizes the critical importance of continuous, state-of-the-art monitoring of Io's volcanic activity to achieve a more comprehensive understanding. We hope these observations and trends can inform future studies when more examples of outbursts enable more rigorous analyses.

# 7 Acknowledgements

Special thanks to John R. Spencer for his valuable comments on the manuscript. Part of this work was carried out at the Jet Propulsion Laboratory, California Institute of Technology, under contract with NASA.